\documentclass{aa}  
\usepackage{hyperref}
\usepackage{microtype}
\usepackage{graphicx}
\usepackage{txfonts}
\usepackage{lipsum}
\usepackage{subcaption}         
\usepackage{lscape}             
\usepackage{placeins}           
\usepackage{float}
\usepackage{caption}

\begin{document}

   \title{Planetesimal formation facilitated by streaming instability\\
in weak pressure bumps}


%
%
%

   \author{Magnus S. Sørensen-Taylor\inst{1,2,3}\thanks{Corresponding author: m.s.sorensentaylor@uva.nl}\and                    Anders Johansen\inst{1,4} \and Troels Haugbølle\inst{2}}

   \institute{Center for Star and Planet Formation, Globe Institute, University of Copenhagen, Øster Voldgade 5-7, 1350 Copenhagen, Denmark
   \and Niels Bohr Institute, University of Copenhagen, Jagtvej 155A, DK-2200 Copenhagen, Denmark
   \and Anton Pannekoek Institute for Astronomy, University of Amsterdam, Science Park 904, 1098 XH Amsterdam, The Netherlands
   \and Division of Astrophysics, Department of Physics, Lund University, Box 118, 22100 Lund, Sweden 
}

   \date{Received ...}

 
  \abstract
    {Planetesimal formation via the streaming instability of small pebbles in protoplanetary discs requires enhanced solid concentrations relative to the solar metallicity, $Z\simeq 0.01$. We investigate here whether pile-ups in weak axisymmetric pressure bumps are sufficient to trigger strong particle concentrations via the streaming instability. Using high-resolution 2D shearing box simulations with millimetre-to-centimetre pebbles, we explore the behaviour of the streaming instability in the presence of a non-reinforced pressure bump. We find that even very weak bumps, with gas density amplitudes as low as $A = 0.04$ relative to the background, produce dense particle filaments via the streaming instability for all tested Stokes numbers at solar metallicity in the inner disc. Outer disc regions require slightly stronger bumps ($A \geq 0.14$) to form filaments at a solar metallicity, though the necessary bump amplitude is substantially lowered when the metallicity is increased to $Z = 0.02$. Our results suggest that weak, non-reinforced pressure bumps can act as focal points for planetesimal formation via the streaming instability in small pebbles, in contrast to previous studies which used reinforcement to maintain the pressure bump. Additionally, we introduce a pressure‑bump‑dependent clumping criterion, $(Z/\chi)_\mathrm{crit} \approx 0.3$, where $\chi = \Pi_\mathrm{min}^2/\Pi_0$ reduces to the background pressure gradient $\Pi_0$ in the absence of a pressure bump. This criterion encapsulates the scale of solid pile-ups at lower pressure gradients and accurately predicts the onset of strong clumping based on the results of our simulations. With pressure bump signatures being common features of observed young discs and magnetohydrodynamical simulations, the results of our 2D simulations imply that weak pressure bumps may be major cradles for planetesimal formation in protoplanetary discs.}


   \maketitle

\makeatletter
\@fleqnfalse
\makeatother
\nolinenumbers
\section{Introduction}
The earliest stages of planet formation are governed by the dynamics and evolution of solid particles in protoplanetary discs. Individual dust particles initially grow from micrometre-sized grains to millimetre- and centimetre-sized pebbles via collisional sticking and mass transfer, though further growth is inhibited by fragmentation or bouncing (\citealp{Blum08}; \citealp{Guttler10}; \citealp{Zsom10}; \citealp{Birnstiel12}; \citealp{Dominik24}). Forming kilometre-scale planetesimals likely requires gravitational collapse of dense dust layers (\citealp{Goldreich73}; \citealp{Johansen14}), which demands that particle densities reach values comparable to the Roche density, $\rho_\mathrm{R}$. Such high densities are difficult to achieve, as solids are removed from the disc through radial drift (\citealp{Weidenschilling77}; \citealp{Laibe12}; \citealp{Birnstiel12}) whilst turbulence counteracts gravitational sedimentation \citep{Weidenschilling80}. 

The streaming instability (SI) offers a mechanism to circumvent these formative barriers. The SI arises from mutual aerodynamic drag forces between gas and radially drifting solid particles and, in its non-linear phase, causes solids to spontaneously concentrate into axisymmetric filaments (\citealp{Johansen07}; \citealp{Johansen09}). The evolution of the SI is governed by the solid-to-gas volume density ratio,

\begin{equation}
    \epsilon =\rho_\mathrm{p}/\rho_\mathrm{g},
\end{equation}

\noindent where $\rho_\mathrm{p}$ and $\rho_\mathrm{g}$ are the volume densities of solid particles and gas, respectively, and by the particle size through the Stokes number, 

\begin{equation}
    \tau_\mathrm{s} = t_\mathrm{stop} \Omega_\mathrm{K},
\end{equation}

\noindent where $t_\mathrm{stop}$ is the aerodynamical stopping time of solid particles and $\Omega_\mathrm{K}$ is the Keplerian orbital frequency. Rapid growth of the SI occurs for $\epsilon \geq 1$ \citep{Youdin05}, producing clumps that may become gravitationally unstable (\citealp{Simon16}; \citealp{Schafer17}; \citealp{Li19}). Lower solid-to-gas ratios exhibit much lower growth rates, particularly if the solids have a broad size distribution (\citealp{Bai10b}; \citealp{Krapp19}; \citealp{Yang21}; \citealp{Schaffer21}; \citealp{Zhu21}). High-resolution simulations have mapped the disc conditions required to promote strong SI-induced clumping (\citealp{Carrera15}; \citealp{Yang17}; \citealp{Li21}). These studies show that clumping-favourable regimes correspond to a wide range of Stokes numbers ($0.001 \leq \tau_\mathrm{s} \leq 1$) and a sufficiently high solid-to-gas column density ratio, hereafter referred to as the ''metallicity'', $Z=\Sigma_\mathrm{p} / \Sigma_\mathrm{g}$. Turbulence has been found to regulate the evolution of the SI (\citealp{Gole20}; \citealp{Lim24}), though magnetohydrodynamical simulations have shown that the SI can evolve and co-exist with a variety of other instabilities (\citealp{Yang18}; \citealp{Schafer20}; \citealp{Eriksson26}). The radial pressure gradient also influences SI thresholds \citep{Bai10a}, with suppression at large pressure gradients being mitigated by an appropriate scaling of the metallicity threshold \citep{Sekiya18}.  

Both observations and protoplanetary disc models show that the solid reservoir in protoplanetary discs is depleted over a few million years (\citealp{Pinilla12}; \citealp{Ansdell16}; \citealp{Appelgren23}; \citealp{Gurrutxaga24}). Therefore, additional mechanisms to concentrate solid particles may be needed to reach metallicities that permit filament formation by the SI. ALMA observations \citep{ALMA15} have revealed that protoplanetary discs contain prominent axisymmetric dust rings at millimetre wavelengths \citep{Andrews18}, widely interpreted as signatures of pressure bumps \citep{Pinilla12}. In such regions, the orbital velocity of the gas approaches or exceeds the Keplerian speed, diminishing the azimuthal headwind felt by solids and enabling radial trapping of solid particles. Pressure bumps have previously been found to be generated by magnetorotational turbulence (\citealp{Johansen09}; \citealp{Eriksson26}), MRI dead zones (\citealp{Kretke09}; \citealp{Flock15}; \citealp{Iwasaki24}), and magnetised disc winds (\citealp{Bethune17}; \citealp{Riols19}), suggesting multiple pathways for pressure bumps to arise. Gaps carved by planets are also suggested to be the cause of axisymmetric pressure maxima (\citealp{Rosotti16}; \citealp{Fedele17}; \citealp{Zhang18}; \citealp{Eriksson21}; \citealp{Stadler22}). 

Planetesimal formation at pressure bumps may be responsible for the marginal optical thickness observed in dust rings \citep{Stammler19}, whilst 3D hydrodynamical simulations have also established favourable conditions for planetesimal formation via the SI for centimetre-sized particles at weakly-reinforced pressure bumps \citep{Carrera21}. Subsequent hydrodynamical simulations showed that the SI was unable to form planetesimals for millimetre-sized particles at reinforced bumps \citep{Carrera22}, suggesting that dust aggregates must grow to much larger sizes before pressure bumps can provide favourable conditions for planetesimals to form. However, applying a reinforcement scheme to uphold the gas density profile could potentially dampen the mutual gas-particle interactions of the pure SI. Such damping may become significant for smaller particles and short reinforcement timescales, as the gas can be modified faster by the reinforcement mechanism than by the particle feedback. As a consequence, a thorough study of the pure SI in non-reinforced pressure bumps is warranted. 

In this work, we use high-resolution 2D hydrodynamical simulations of the SI to examine how radial trapping of millimetre-to-centimetre sized pebbles at a pressure bump may assist SI-driven clumping. We impose a Gaussian enhancement of the gas density relative to the background gradient and explore variations in pressure bump amplitude, background pressure support, particle Stokes number, and metallicity. The density enhancement is not reinforced, allowing the SI to develop naturally within the pressure bump. Solid particle densities are compared to the characteristic Roche density when assessing whether filaments produced by the SI can collapse gravitationally to form planetesimals, as self-gravity is not explicitly included in our simulations.    

This paper is organised as follows. In Section \ref{methods} we describe the local disc properties and the setup of our numerical simulations. In Section \ref{results} we present the results of how the SI evolves at various pressure bump configurations in 2D simulations. In Section \ref{discussion} we discuss a new parametrisation of the clumping criterion that also considers gas pressure perturbations, with subsequent discussion about the implications of our chosen setups in Section \ref{discussion2}. Finally, Section \ref{conclusion} summarises our findings and provides context for future work. Additional discussion about pressure bump deformation and the implementation of extended pressure bumps is provided in Appendices \ref{sec:deformapp} and \ref{sec:robust}, respectively.
\section{Methods}\label{methods}
 We simulate the dynamics of gas and solid particles in a 2D shearing box approximation (\citealp{Goldreich65}; \citealp{Goldreich78}; \citealp{Hawley95}), with a reference distance from the central star, $r_0$, and a reference Keplerian orbital frequency, $\Omega_0$. This allows for the dynamics in the corotating frame to be described using Cartesian coordinates. We assume the box to represent a locally thin disc, where the gas scale height, $H=c_\mathrm{s} /\Omega_\mathrm{K}$, is much smaller than the distance from the central star, $r$.

We use the \textsc{Pencil Code}\footnote{More details at https://pencil-code.nordita.org/} to solve the hydrodynamics in the shearing box \citep{Pencil21}, employing a grid-based description for the gas and a Lagrangian superparticle description for solid particles (\citealp{Wisdom88}; \citealp{Youdin07}). The equations of motion for gas and solids are solved in an Eulerian grid consisting of $N_x \times N_z = 9216 \times 256$ grid cells, where $x$ is the radial coordinate and $z$ is the vertical coordinate. The grid is initialised with physical dimensions $L_x \times L_z = 7.2H \times0.2H$, giving us a grid resolution of $1280 H^{-1}$, and we use one superparticle per cell. Vertical sedimentation subsequently increases the superparticle number density near the midplane, placing the midplane layer well within the regime where SI growth rates of monodisperse and polydisperse particle size distributions converge (\citealp{Youdin07}; \citealp{Schaffer21}). Each of our simulations uses a monodisperse distribution of pebble sizes whilst disregarding aspects of dust evolution, though dust coagulation would most likely synergise positively with the SI by producing particle sizes favourable to clumping (\citealp{Ho24}; \citealp{Carrera25}; \citealp{Vallucci-Goy26}). We also neglect any contribution from external turbulence, but note that planetesimal formation via the SI
has been shown to persist in turbulence generated by the magnetorotational instability \citep{Eriksson26}. Our boundary conditions are shear-periodic across the radial domain and periodic across the vertical domain, hence prohibiting any mass loss across the shearing box over the simulated timeframe. Periodic vertical boundaries can, in some cases, introduce numerical artefacts when outflowing material re‑enters the domain and adds artificial stirring to the midplane layer. Such effects can be alleviated by using large domains, or by adopting outflow boundary conditions (\citealp{Li18}; \citealp{Li21}; \citealp{Lim26}). We leave an exploration of the effect of different boundary conditions on the evolution of the SI in weak pressure bumps to future work.

\subsection{Role of pressure support on solid transport}
We now consider how the general dynamics between gas and solid particles evolve in a local patch of a protoplanetary disc. The gas experiences a radially varying pressure support which sets the azimuthal headwind felt by solids, denoted by the dimensionless pressure support term,  

\begin{equation}\label{eq:eta}
\eta = -\frac{1}{2}\left(\frac{c_\mathrm{s}}{v_\mathrm{K}}\right)^2
\frac{\partial \ln P}{\partial \ln r} \;.
\end{equation}

\noindent Here $c_\mathrm{s}$ is the local sound speed, $v_\mathrm{K}$ is the local Keplerian orbital velocity, and ${\partial \ln P}/{\partial \ln r}$ is the global radial pressure gradient \citep{Nakagawa86}. A negative pressure gradient ($\eta > 0$) forces gas to orbit at a sub-Keplerian velocity, $\Delta v = \eta v_\mathrm{K}$, relative to solids orbiting at the Keplerian velocity, $v_\mathrm{K}$. The motion of solids, regulated by the aerodynamic drag instilled by the orbital velocity differential with the gas, is governed directly by the radial pressure gradient through the dimensionless pressure gradient parameter, $\Pi$, also denoted as the headwind parameter,

\begin{equation}\label{eq:Pi}
\Pi = \frac{\eta v_\mathrm{K}}{c_\mathrm{s}} = -\frac{1}{2}\left(\frac{H}{r}\right)
\frac{\partial \ln P}{\partial \ln r} \;,
\end{equation}

\noindent The extent of the aerodynamic drag experienced by solid particles is dependent on particle size. Up to roughly metre-sizes, the drag-coupling between gas and solids is quantified by a particle stopping time in the Epstein regime \citep{Epstein24},

\begin{equation}\label{eq:t_stop}
    t_{\mathrm{stop}} = \frac{\rho_\mathrm{s} a}{\rho_\mathrm{g} c_\mathrm{s}},
\end{equation}

\noindent where $a$ and $\rho_\mathrm{s}$ are the radius and material density of solid particles. The azimuthal headwind experienced by solid particles instils an angular momentum loss in solids. This incurs a speed, 

\begin{equation}\label{eq:w_x}
    v_x \approx -2 \tau_\mathrm{s}\Delta v ,
\end{equation}

\noindent of the radial drift of small pebbles (\citealp{Adachi76}; \citealp{Weidenschilling77}). The highest radial drift speeds are obtained for solid particles of Stokes number $\tau_\mathrm{s} = 1$, with smaller Stokes numbers incurring lower radial drift speeds. Solid particles also sediment in the midplane of the disc due to vertical gravity, though this occurs over a much shorter timeframe. The time required for solids to sediment fully is given by the settling timescale, $t_{\mathrm{settle}} = 1/(\Omega_{\mathrm{K}}\tau_\mathrm{s})$ \citep{Dubrulle95}.
 
\subsection{Pressure bump setup}\label{sec:pbsetup}
Using an isothermal equation of state, a pressure change to the gas can be induced by adding a perturbation in the gas density relative to the background gradient. In geostrophic balance, this also induces a perturbation in the azimuthal velocity of the gas surrounding the bump. We define the perturbation in the radial gas density profile relative to the background gradient as a Gaussian profile with amplitude $A$. In vertical hydrostatic equilibrium, the perturbed gas density profile yields

\begin{equation}\label{eq:rho_g}
     \rho_\mathrm{g}(x,z) = \rho_0\left[1+A \mathrm{e}^{-x^2/2w^2} \right]\mathrm{e}^{-z^2/2H^2} ,
\end{equation}

\noindent where $\rho_0$ is the background midplane gas density, and $w$ is the pressure bump width. The resulting azimuthal gas velocity profile in geostrophic balance is independent of the $z$-direction, and is given by

\begin{equation}\label{eq:vyg}
    u_{y}(x) = \frac{c_\mathrm{s}^2}{2 \Omega_0}\left( \frac{\partial \ln\rho_0}{\partial x}-\frac{Ax\mathrm{e}^{-x^2/2w^2}}{w^2[1+A\mathrm{e}^{-x^2/2w^2}]} \right),
\end{equation}

\noindent This profile describes both the perturbed azimuthal gas velocity and the background Keplerian-shear-subtracted velocity profile, $u_{y,0}= -\Pi_0 c_\mathrm{s} =({c_\mathrm{s}^2}/{2 \Omega_0})({\partial \ln\rho_0}/{\partial x})$, where $\Pi_0$ is the background pressure support. Since $\Pi = -u_y/c_\mathrm{s}$, Equation \ref{eq:vyg} indirectly describes the radial variation in the pressure gradient parameter across the shearing box. The local maximum in the azimuthal gas velocity corresponds to a minimum headwind, $\Pi_\mathrm{min}$, at which the radial drift of solid particles is slowest (\citealp{Weidenschilling77}; \citealp{Nakagawa86}).

   \begin{figure}[t!]
   \centering
   \includegraphics[width=\hsize]{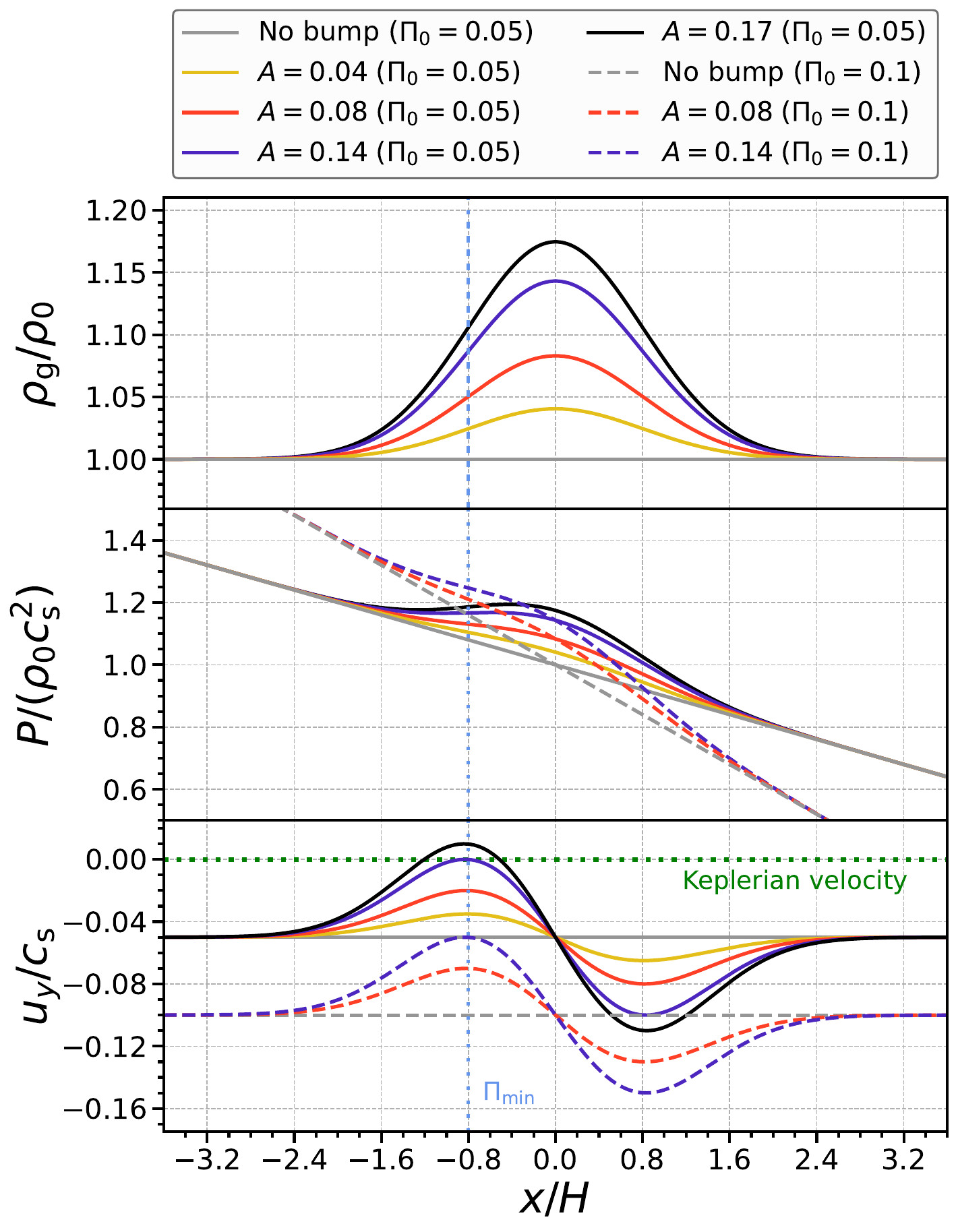}
      \caption{Initial radial profiles of the gas density relative to the background gradient (top), full gas pressure (middle), and Keplerian-shear-subtracted azimuthal gas velocity (bottom) using different bump amplitudes, $A$. The pressure and the Keplerian-shear-subtracted azimuthal gas velocity have two different profiles: one corresponding to lower background pressure support, $\Pi_0 = 0.05$ (solid lines), and the other corresponding to higher background pressure support, $\Pi_0 = 0.1$ (dashed lines).}
         \label{fig:initprofs}
   \end{figure}

\subsection{Initial conditions}
We initialise our simulations with the pressure bump setup outlined in Section \ref{sec:pbsetup}. All setups have an equal pressure bump width, $w=0.8H$, to ensure that the difference in shear does not trigger the Rossby wave instability (RWI; \citealp{Lovelace99}), after which we vary bump amplitude, particle Stokes number, metallicity, and background pressure support independently. We use fairly weak pressure bump amplitudes, ranging from $A=0.04$ to $A=0.17$ across our simulations. The corresponding radial profiles of the gas density relative to the background density gradient, $ \rho_\mathrm{g}(x)$, pressure, $P(x)$, and Keplerian-shear-subtracted azimuthal gas velocity, $u_y(x)$, are shown in Figure \ref{fig:initprofs}. Employing these amplitudes and our chosen bump width, $w$, puts the location of the minimum headwind at $x(\Pi_\mathrm{min}) \approx -0.8H$ for all bump amplitudes\footnote{The location of the minimum headwind, $\Pi_{\rm{min}}$, shifts slightly as the bump amplitude changes, with higher amplitudes moving $\Pi_{\rm{min}}$ closer to the bump center. However, these differences are very minor for our range of amplitudes.}. Our Stokes numbers correspond to the millimetre-to-centimetre pebble size range, evaluating cases with $\tau_\mathrm{s} = 0.01$, $\tau_\mathrm{s} = 0.03$, and $\tau_\mathrm{s} = 0.1$ individually in order to adequately explore the overlap between solid particle sizes observed in protoplanetary discs which also display significant clumping from the SI. We use metallicities $Z=0.01$, corresponding to the approximate solar metallicity, and $Z=0.02$, obtainable e.g. via photoevaporation in the disc \citep{Carrera17} or by ice pile-up at the water ice line (\citealp{Drazkowska17}; \citealp{Schoonenberg17}; \citealp{Ros24}). Finally, we explore the effect of the background pressure support, implementing pressure bumps in regions characterised by background pressure gradient parameters $\Pi_0 = 0.05$ and $\Pi_0 = 0.1$. A total of 8 different combinations of $\Pi_0$ and $A$ are chosen, giving us a wide variety of prescriptions for the background pressure gradient minimum, described by $\Pi_\mathrm{min}$ (see Table \ref{tab:inits2}). The specifications of all simulation setups are given in Table \ref{tab:inits}. 

We assume a disc model with a gas column density profile scaling as $\Sigma_\mathrm{g}(r) \propto r^{-1}$ and a stellar irradiation temperature profile for a solar-mass, solar-luminosity star, $T(r) \propto r^{-3/7}$ (\citealp{Garaud07}; \citealp{Oka11}; \citealp{Ida16}). The gas scale height scales as $H(r) \propto r^{9/7}$, yielding a radial profile in the background pressure gradient parameter scaling as $\Pi_0(r) \propto r^{2/7}$. Using typical constants in the gas density and temperature profiles, the final prescription for the radial profile of the global background pressure gradient parameter is given by

\begin{equation}
    \Pi_0 = 0.034 \left( \frac{r}{1 \; \mathrm{AU}} \right)^{2/7}.
\end{equation}

\noindent Here $\Pi_0 = 0.05$ corresponds to $r_0 \approx 4\,\mathrm{AU}$, while $\Pi_0 = 0.1$ corresponds to $r_0 \approx 44\,\mathrm{AU}$. The radial temperature profile not only sets the background pressure gradient but also determines the density threshold at which solid particle clumps become self-gravitating. This threshold is given by the Roche limit, expressed as the ratio between the Roche density, $\rho_\mathrm{R}$, and the background midplane gas density, $\rho_0$,
\begin{equation}
    \frac{\rho_\mathrm{R}}{\rho_0} = \frac{9}{\Gamma}.
\end{equation}
\noindent Here $\Gamma$ is the dimensionless self-gravity parameter, which depends on the disc temperature $T$, the viscous parameter $\alpha$, and the stellar accretion rate $\dot{M}_\star$ \citep{Johansen26} as
\begin{equation}
    \Gamma \approx 0.11 
    \left( \frac{\dot{M}_\star}{10^{-7}\,\mathrm{M}_\odot\,\mathrm{yr}^{-1}} \right)
    \left( \frac{\alpha}{10^{-2}} \right)^{-1}
    \left( \frac{T}{100\,\mathrm{K}} \right)^{-3/2}.
\end{equation}
\noindent Adopting typical values for observed young stellar object systems, $\alpha = 10^{-3}$ and $\dot{M}_\star = 7\times10^{-9}\,\mathrm{M}_\odot\,\mathrm{yr}^{-1}$ (\citealp{Alcala17}; \citealp{Villenave25}), we obtain Roche densities of $\rho_{\mathrm{R},4} \approx 88\rho_0$ and $\rho_{\mathrm{R},44} \approx 19\rho_0$ for $r_0 \approx 4 \; \mathrm{AU}$ and $r_0 \approx 44 \; \mathrm{AU}$, respectively.

\begin{table}[t!]
\centering
\small
\begin{tabular*}{\columnwidth}{@{\extracolsep{\fill}}@{\hspace{28pt}}c c c@{\hspace{28pt}}}
\hline\hline
\noalign{\smallskip}
$\Pi_0$ & $A$ & $\Pi_{\min}$\\
\noalign{\smallskip}
\hline
0.05 & 0 & 0.05 \\
0.05 & 0.04 & 0.035 \\
0.05 & 0.08 & 0.02 \\
0.05 & 0.14 & 0 \\
0.05 & 0.17 & -0.01 \\
0.1  & 0 & 0.1 \\
0.1  & 0.08 & 0.07 \\
0.1  & 0.14 & 0.05 \\
\hline
\end{tabular*}
\caption{Values of the minimum headwind $\Pi_\mathrm{min}$, for different background pressure gradient parameters, $\Pi_0$, and bump amplitudes, $A$, used in our simulations.}
\label{tab:inits2}
\end{table}

\begin{table}[t!]
\centering
\small
\begin{tabular*}{\columnwidth}{@{\extracolsep{\fill}}@{\hspace{28pt}}c c c@{\hspace{28pt}}}
 \hline \hline
 \noalign{\smallskip}
$A$&$\tau_\mathrm{s}$&$Z$\\
 \noalign{\smallskip}
 \hline
 \noalign{\smallskip}
\multicolumn{3}{c}{ $\Pi_0 = 0.05$} \\
\noalign{\smallskip}
\hline
0, 0.04, 0.08, 0.14, 0.17 & 0.01 & 0.01 \\
0, 0.04, 0.08, 0.14, 0.17 & 0.03 & 0.01 \\
0, 0.04, 0.08, 0.14, 0.17 & 0.1 & 0.01 \\
\hline
\noalign{\smallskip}
\multicolumn{3}{c}{ $\Pi_0 = 0.1$} \\
\noalign{\smallskip}
\hline
0, 0.08, 0.14 & 0.01 & 0.01 \\
0, 0.08, 0.14 & 0.01 & 0.02 \\
0, 0.08, 0.14 & 0.03 & 0.01 \\
0, 0.08, 0.14 & 0.03 & 0.02 \\
0, 0.08, 0.14 & 0.1 & 0.01 \\
0, 0.08, 0.14 & 0.1 & 0.02 \\
\hline
\end{tabular*}
\caption{Specifications of all simulation runs, split into two groups based on the value of the background pressure parameter, $\Pi_0$. Note that each bump amplitude value constitutes its own simulation run. All setups have a bump width of $w=0.8H$ and a particle-per-cell count of $n_\mathrm{p}\approx 1$ across the whole 2D box. Each simulation is run until significant particle clumping takes effect, or until clear signs of stagnation are exhibited.}
\label{tab:inits}
\end{table}

\section{Results}\label{results}
We now present the overarching set of results using different input parameters. The most prominent pattern exhibited in our simulations is that weak pressure bumps are sufficient to induce strong clumping via the SI for pebbles in the millimetre-to-centimetre size range, both in the inner and outer disc. In the inner disc region where $\Pi_0 \sim 0.05$, amplitudes down to $A=0.04$ induce clumping for particles of Stokes numbers $\tau_\mathrm{s}=0.01$, $\tau_\mathrm{s}=0.03$, and $\tau_\mathrm{s} = 0.1$, with a solar metallicity, $Z=0.01$. In the outer disc region where $\Pi_0 \sim 0.1$, all three tested Stokes numbers were only found to induce clumping for a bump of amplitude $A\geq 0.14$ when using a solar metallicity. However, clumping limits were clearly bypassed when using a super-solar metallicity, $Z=0.02$ for bump amplitudes down to $A=0.08$.

    \begin{figure*}
        \centering
        \includegraphics[width=1\textwidth]{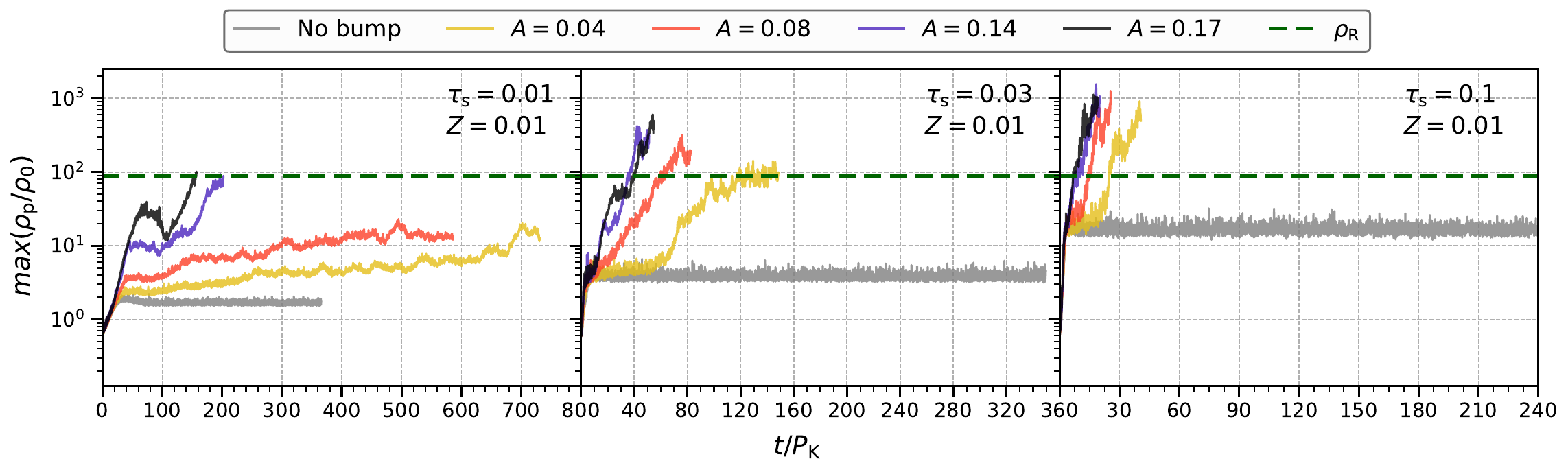}
        \caption{The measured maximum in the local solid particle density relative to the background midplane gas density as a function of time, for simulations using a background pressure parameter of $\Pi_0 = 0.05$ and an initial metallicity of $Z = 0.01$. Each individual panel shows the evolution, in units of Keplerian orbits, $P_\mathrm{K}$, of the maximum solid particle density for simulations using different Stokes numbers, $\tau_\mathrm{s}$. Individual runs are represented by a colour unique to the pressure bump amplitude, $A$, including simulations without a pressure bump.} 
        \label{fig:grouped}%
    \end{figure*}

    \begin{figure*}
        \centering
        \includegraphics[width=1\textwidth]{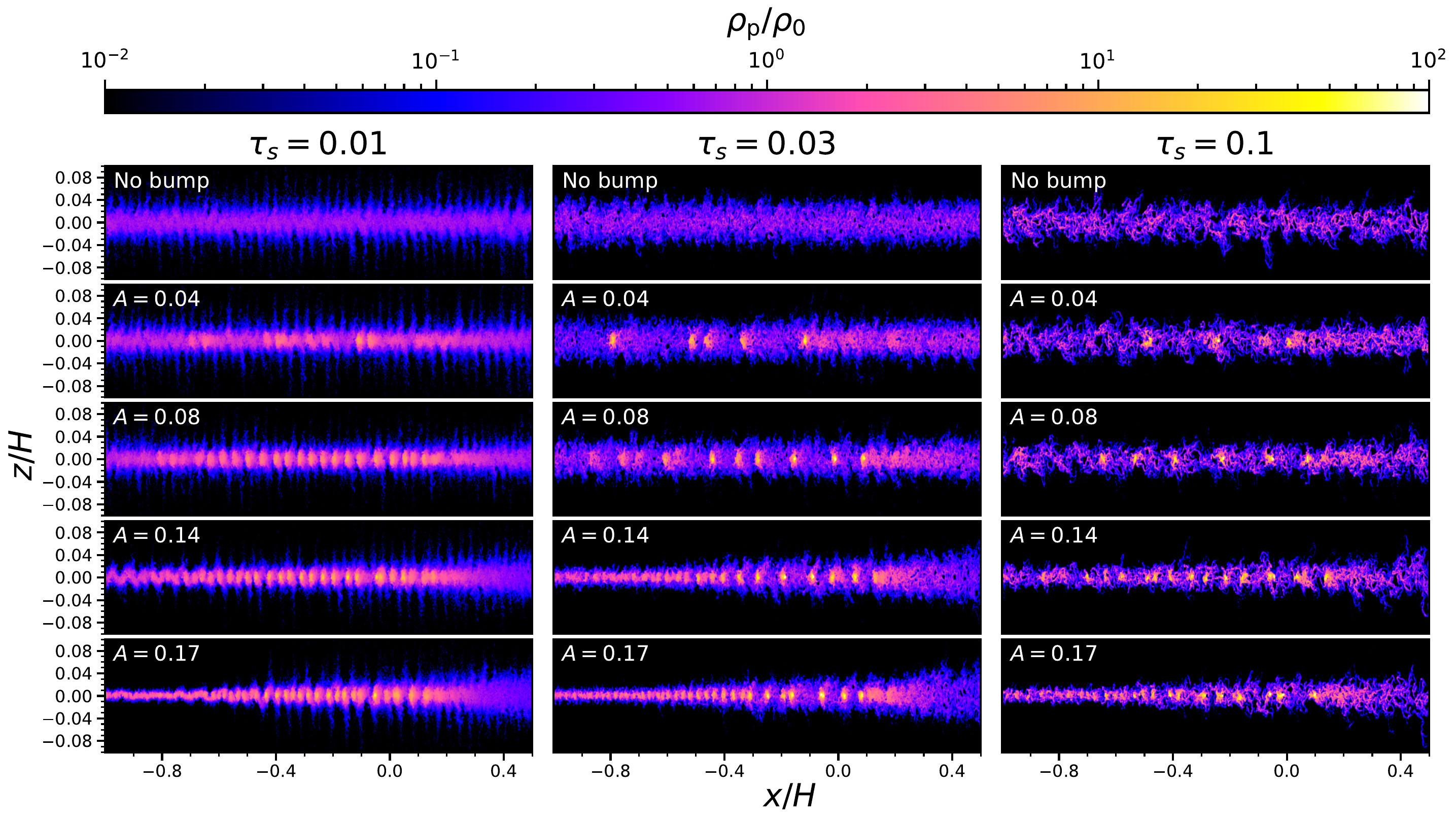}
        \caption{Final simulation snapshots of the solid particle density relative to the background midplane gas density, $\rho_\mathrm{p}/\rho_0$ (represented by the colour scale), for runs with a background pressure gradient parameter of $\Pi_0 = 0.05$ and an initial metallicity of $Z=0.01$. Each column represents simulations using one particular Stokes number, $\tau_\mathrm{s}$, and each row represents simulations using one particular bump amplitude, $A$. We show a portion of the radial domain, $-H\leq x \leq 0.5H$, corresponding to where the solid particle density enhancement from the pressure maximum is most prominent in setups with a bump present.} 
        \label{fig:rhop_matrix}%
    \end{figure*}

\subsection{Bumps in the inner disc}\label{sec:innerdisc} 
\begin{figure}[t!]
   \centering
   \includegraphics[width=\hsize]{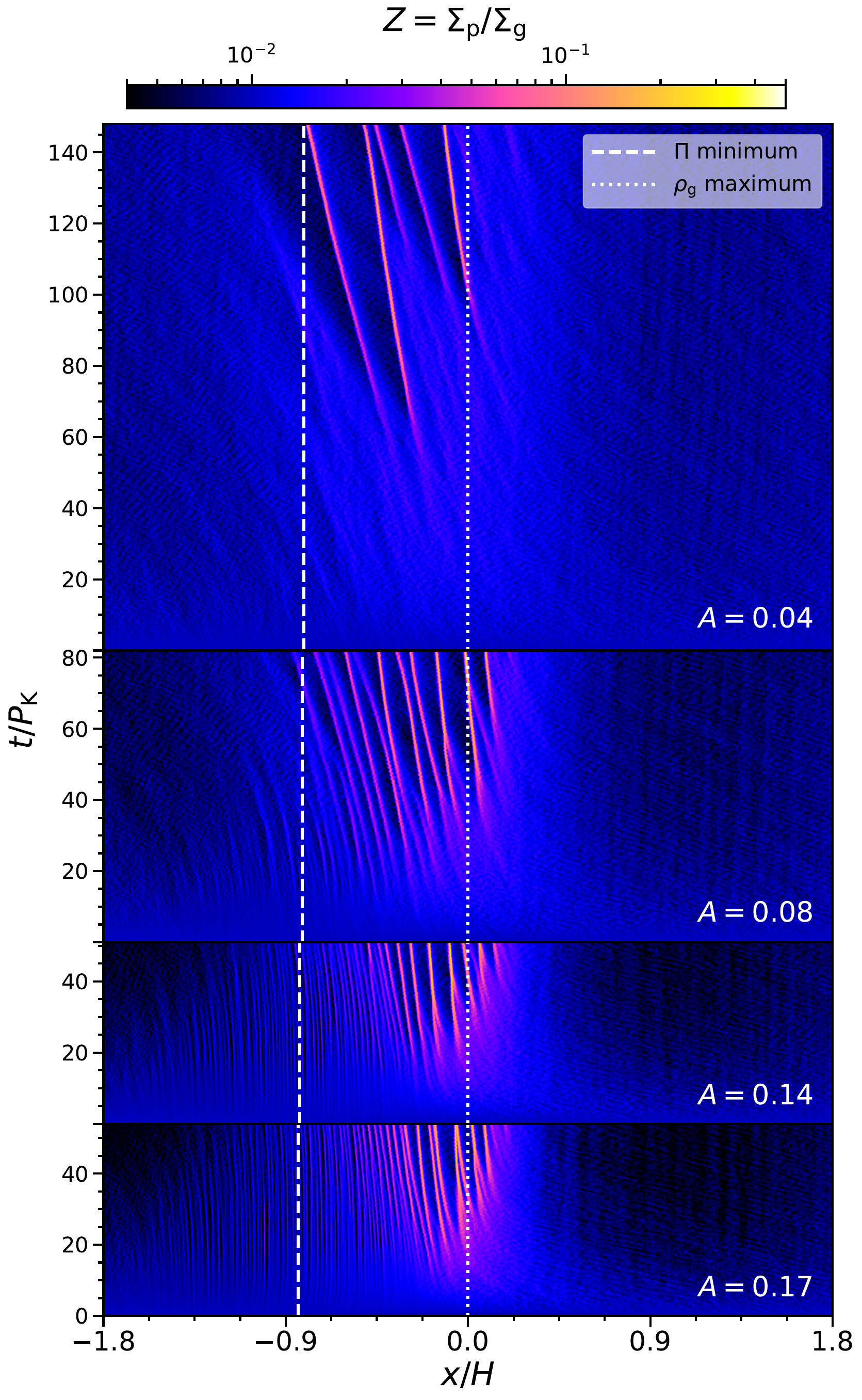}
      \caption{Time evolution of the metallicity, $Z=\Sigma_\mathrm{p}/\Sigma_\mathrm{g}$ (represented by the colour scale), for solid particles of Stokes number $\tau_\mathrm{s} = 0.03$, a background pressure parameter of $\Pi_0 = 0.05$, an initial metallicity of $Z=0.01$, and pressure bump amplitudes $A=[0.04, 0.08, 0.14, 0.17]$. Each panel shows a portion of the radial domain encompassing both the headwind minimum and the gas density maximum.}
         \label{fig:spacetimeplot}
   \end{figure}

Runaway particle concentration via the SI in weak pressure bumps is particularly prominent in inner disc regions where the background pressure support is smaller. Our simulations showcase that the lower overall azimuthal headwind induced by a smaller background pressure support ($\Pi_0 = 0.05$) allows pressure bumps of amplitudes down to the lowest value used in our investigation ($A=0.04$, corresponding to $\Pi_\mathrm{min}=0.035$) to induce strong particle clumping for particles of Stokes numbers $\tau_\mathrm{s} = 0.01$, $\tau_\mathrm{s} = 0.03$, and $\tau_\mathrm{s} = 0.1$. Particle clumping at the pressure bump is expedited by a higher bump amplitude, occurring fastest for particles of Stokes number $\tau_\mathrm{s} = 0.1$ and more gradually for Stokes numbers $\tau_\mathrm{s} = 0.03$ and $\tau_\mathrm{s} = 0.01$, as shown in Figure \ref{fig:grouped}. Figure \ref{fig:rhop_matrix} shows the solid particle volume densities for all simulations at their final snapshots. Simulations with a pressure bump exhibit the formation of dense particle filaments where rapid solid density growth occurs, with these filaments generally forming on the starward side ($x<0H$) of the initial gas density bump. Radial spacing between filaments also increases for larger Stokes numbers and higher values of $\Pi_\mathrm{min}$, corresponding to the resonant wavelength of the fastest SI growing modes, $\lambda_{\mathrm{SI},x} \approx 4 \pi \tau_\mathrm{s}\Pi H$ \citep{Squire20}. A notable detail from the runs using higher bump amplitudes ($A=0.14$ and $A=0.17$, corresponding to $\Pi_\mathrm{min}=0$ and $\Pi_\mathrm{min}=-0.01$, respectively) is that filaments satisfying $\rho_\mathrm{p}  \geq \rho_\mathrm{R}$ tend not to form at $x(\Pi_\mathrm{min})$, but rather closer to the center of the pressure enhancement, where the radial rate of change in $\Pi(x)$ is largest (see Figure \ref{fig:spacetimeplot}). This likely reflects a brief pile‑up where solids decelerate most strongly, momentarily promoting particle concentration into filaments before a steady state in the pebble drift is established. We return to this point in Section \ref{discussion}.

\subsection{Bumps in the outer disc}\label{sec:outerdisc}
\begin{figure*}[t!]
    \centering

    \begin{subfigure}{1.0\textwidth}
        \centering
        \includegraphics[width=\textwidth]{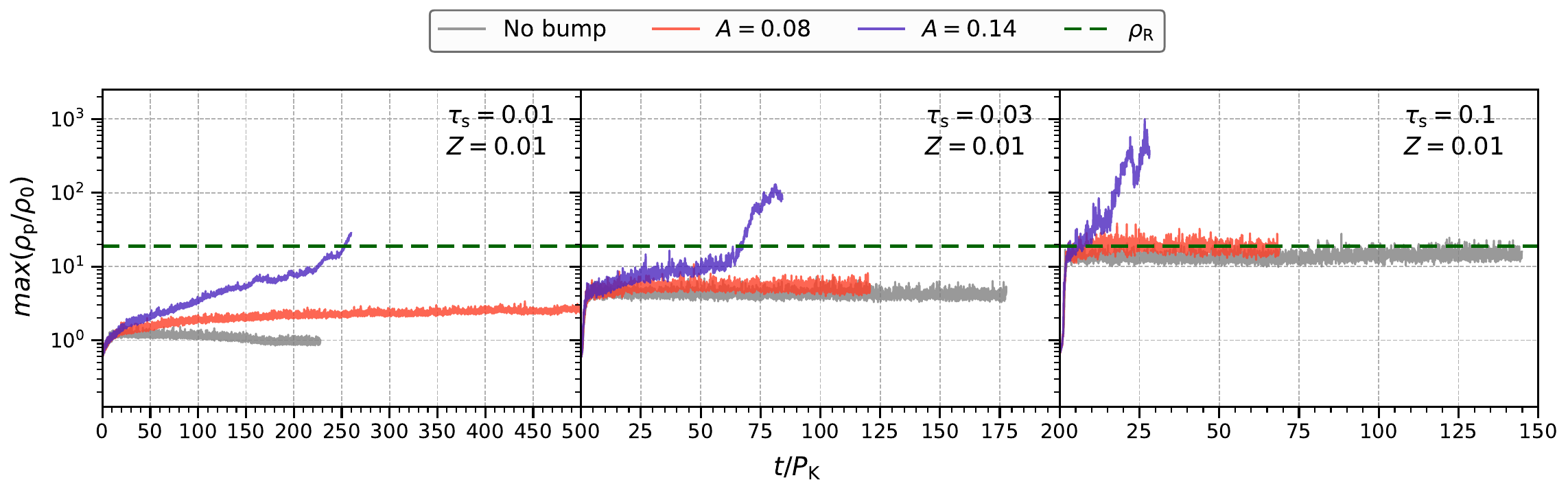}
    \end{subfigure}\label{fig:grouped2}

    \vspace{0.5em}

    \begin{subfigure}{1.0\textwidth}
        \centering
        \includegraphics[width=\textwidth]{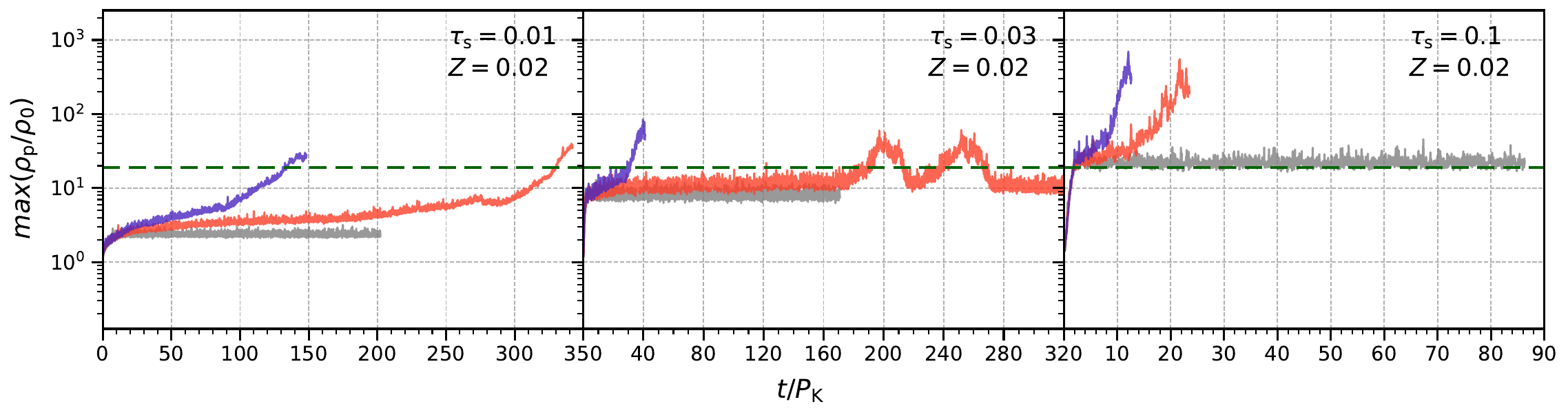}
    \end{subfigure}\label{fig:grouped3}
        \caption{
        The measured maximum in the local solid particle density relative to the background midplane gas density as a function of time, for simulations using a background pressure parameter of $\Pi_0 = 0.1$, Stokes numbers $\tau_\mathrm{s} = [0.01, 0.03, 0.1]$ and pressure bump amplitudes $A=[0, 0.08, 0.14]$. The top panel shows the outcome of simulations using an initial metallicity of $Z=0.01$, while the bottom panel shows the outcome of simulations using an initial metallicity of $Z=0.02$. 
    }
    \label{fig:grouped_Pi0}
\end{figure*}

The growth of the maximum particle density for each bump amplitude using $\Pi_0 = 0.1$ and $Z=0.01$ is shown in the top panel of Figure \ref{fig:grouped_Pi0}, once again comparing pressure bump simulations with bump-absent runs. Whilst our simulations show that very weak pressure bumps in regions of low background pressure support drastically promote streaming unstable conditions by incurring a gradual pile-up of pebbles, they also show that pressure bumps require slightly higher amplitudes to facilitate strong clumping in regions of high background pressure support ($\Pi_0 = 0.1$). This agrees with findings that stronger pressure support of the gas inhibits the formation of particle filaments (\citealp{Bai10a}; \citealp{Sekiya18}). Assessing the two setups with a pressure bump, it is only the bump of higher amplitude ($A=0.14$, corresponding to $\Pi_{\rm{min}}=0.05$) that triggers the formation of dense particle filaments. All three Stokes numbers produce filaments for $\Pi_0=0.1$ when using a bump amplitude of $A=0.14$, though slower compared to equivalent setups using $\Pi_0=0.05$. Dense particle filaments also form less abundantly and closer to $\Pi_\mathrm{min}$ due to the higher overall pebble drift speeds. The particle density growth in these filaments allows for solid particle densities to exceed the Roche limit in all three runs with $A=0.14$, though we note that the Roche density is considerably lower in the $\Pi_0=0.1$ case compared to the $\Pi_0 = 0.05$ case. The bump of lower amplitude ($A=0.08$, corresponding to $\Pi_{\rm{min}}=0.07$) does not form particle filaments for any of the three Stokes numbers. Although there are signs of growth in the solid particle density beyond that of the bump-absent runs, particularly in the $\tau_\mathrm{s} = 0.01$ run, a phase of non-linear particle concentration is never triggered. As $\Pi_{\rm{min}}$ is relatively high for a bump amplitude of $A=0.08$, the limited increase in $Z$ combined with a minimal reduction in $\Pi$ creates a two-fold hindrance to potential clumping. The lower Roche density allows particle concentrations in the $\tau_\mathrm{s}=0.1$ runs to approach the self‑gravitating threshold through vertical sedimentation after entering the weakly-saturated stage of the SI, but collapse remains unlikely because Kelvin–Helmholtz turbulence counteracts additional vertical compression (\citealp{Weidenschilling80}; \citealp{Youdin02}). We also entertain the possibility that a Roche density corresponding to $\rho_{\mathrm{R,44}} \approx 19 \rho_{0}$ is too low even for the conditions in the outer disc region, particularly in the late stages of protoplanetary disc evolution where the cold classical Kuiper belt objects form \citep{Li26}. We therefore emphasise that the presence of any non-linear particle concentration from the SI remains the most important mechanism in determining whether solid particle clumps can become self-gravitating.

\subsection{Runs with super-solar metallicity}
\begin{figure}[t!]
   \centering
   \includegraphics[width=\hsize]{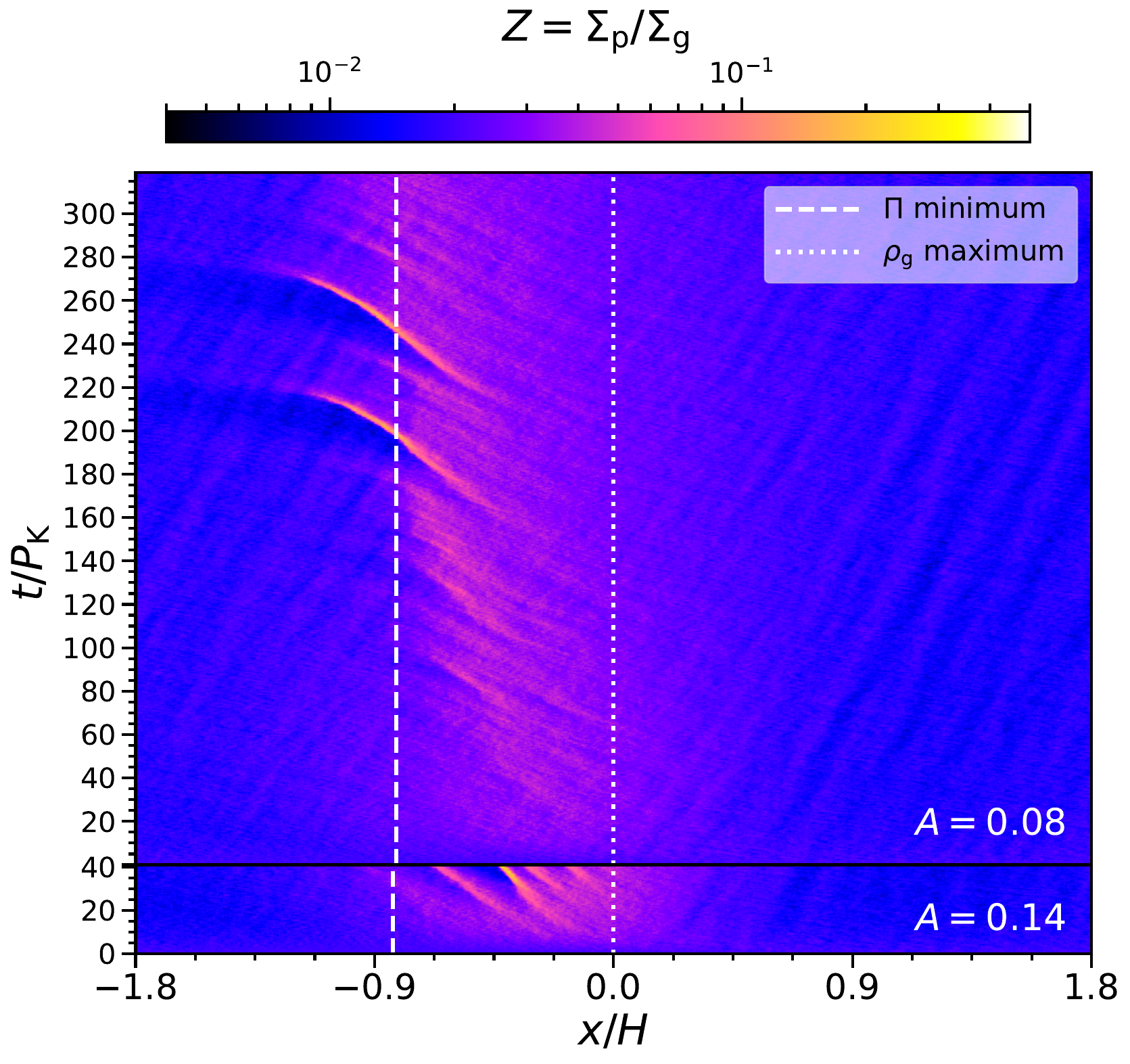}
      \caption{Time evolution of the metallicity, $Z=\Sigma_\mathrm{p}/\Sigma_\mathrm{g}$, for solid particles of Stokes number $\tau_\mathrm{s} = 0.03$, a background pressure parameter of $\Pi_0 = 0.1$, an initial metallicity of $Z=0.02$, and pressure bump amplitudes $A=[0.08, 0.14]$.}
         \label{fig:spacetimeplot5}
   \end{figure}

Although the saturated stage of the SI is strongly controlled by the pressure gradient, it is equally influenced by the metallicity, $Z$, as the metallicity sets the maximum midplane solid density and hence the likelihood of triggering strong clumping (\citealp{Carrera15}; \citealp{Yang17}; \citealp{Li21}). Using the same setups as those detailed in Section \ref{sec:outerdisc}, but with a change to the initial metallicity to $Z=0.02$, we see that bumps down to $A=0.08$ are now also able to facilitate strong particle concentration for Stokes numbers $\tau_\mathrm{s} = 0.01$, $\tau_\mathrm{s}=0.03$, and $\tau_\mathrm{s} = 0.1$ (see the bottom panel of Figure \ref{fig:grouped_Pi0}). As conditions for strong clumping have been found to depend on the initial $Z/\Pi$-ratio in simulations using a uniform pressure gradient \citep{Sekiya18}, this explains why an increase in metallicity allows for a correspondingly lower threshold in the pressure bump amplitude required to initiate strong clumping, rather than relying on those two parameters separately. The exhibited particle concentration allows for the Roche limit to be surpassed in all cases where a bump is present, though particle filaments form very sparsely at bumps of amplitude $A=0.08$. The scarcity of clumps at bumps of amplitude $A=0.08$ using $Z=0.02$ and $\Pi_0=0.1$ is similar to what we observe at bumps of $A=0.14$ using $Z=0.01$ and $\Pi_0=0.1$. These cases may just lie above the threshold for strong clumping at the pressure bump; hence, the radial region harbouring favourable conditions for clumping may be very small. Marginal exceeding of clumping thresholds could certainly be the case for the $\tau_\mathrm{s} = 0.03$ run using $A=0.08$, $\Pi_0 = 0.1$, and $Z=0.02$, in which a single filament only briefly reaches densities above the Roche limit at roughly $t=190P_{\rm{K}}$ and $t=240P_{\rm{K}}$. As the filament drifts, it moves past $x(\Pi_{\rm{min}})$ and disperses shortly afterward (see Figure \ref{fig:spacetimeplot5}). This momentary particle concentration indicates that only a small radial region around $\Pi_\mathrm{min}$ is favourable to clumping via the SI, due to the marginal solid particle pile-up and small local reduction in $\Pi$. However, self-gravity would allow for planetesimals to form on very short timescales before filaments disperse upon leaving the clumping-favourable region, provided the Roche limit is exceeded.

\section{A new pressure-bump-dependent clumping criterion}\label{discussion}
Current SI clumping criteria have been established using a constant value of $\Pi$ (\citealp{Carrera15}; \citealp{Yang17}; \citealp{Li21}), thereby neglecting the effect of a radially-varying pressure gradient. The simulations that we present here encompass a large multitude of different pressure bump simulation setups, which can be used to provide a robust foundation for a clumping criterion based on local variations in the pressure gradient. 

   \begin{figure}[t!]
   \centering
   \includegraphics[width=\hsize]{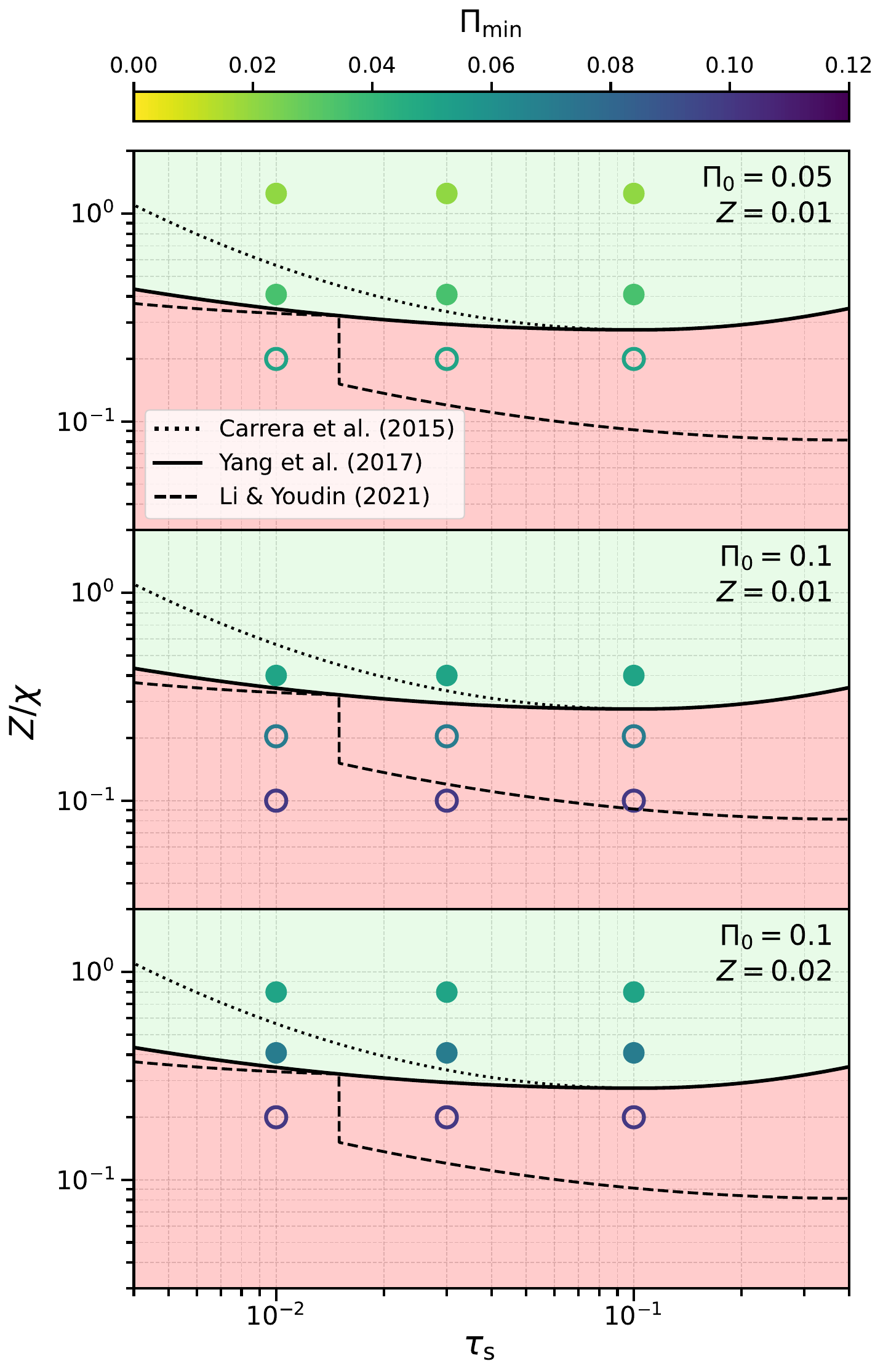}
      \caption{Diagrams of the $Z/\chi$-$\tau_\mathrm{s}$ parameter space, indicating the $Z/\chi$-$\tau_\mathrm{s}$ combinations used in our simulations. These are split into three panels based on the values of $\Pi_0$ and $Z$. Each combination is marked by a circle whose colour corresponds to the value of the minimum headwind, $\Pi_\mathrm{min}$. Runs with visible filament formation are represented by full circles, whilst runs with no visible filament formation are represented by open circles. The points in the $Z/\chi$-$\tau_\mathrm{s}$ parameter space are compared to previous clumping criteria (\citealp{Carrera15}; \citealp{Yang17}; \citealp{Li21}) all using $\chi = 0.05$. The clumping criterion from \cite{Yang17} is highlighted by shading the parameter space according to where their simulations exhibit clumping (green tint) and no clumping (red tint), as our results appear to closely follow this threshold.}
         \label{fig:clumpcrit}
   \end{figure}

We first consider how the pressure-normalised metallicity, $Z/\Pi$ \citep{Sekiya18}, reacts to a region with variations in the pressure support. For an initial uniform distribution of solids relative to the background gas density, a radially-varying pressure gradient will induce a radial profile in the metallicity due to the variation in the radial drift of solids. For a constant mass flux, the resulting difference in advection causes the radial distribution of solids to inversely follow the $\Pi$-profile once the transport of solids comes to a steady state (\citealp{Nakagawa86}; \citealp{Takeuchi02}). As a result, the metallicity is increased in regions where the pebble drift speed is low, and this enhancement can create conditions that favour planetesimal formation driven by the SI. The maximum enhancement in the metallicity given by the steady state solution is simply
\begin{equation}
        Z_{\rm{max}}= Z\left(\frac{\Pi_0}{\Pi_\mathrm{min}}\right)
\end{equation}
and the pressure-normalised metallicity at the location of minimum headwind is subsequently given by 
\begin{equation}
    \frac{Z_{\rm{max}}}{\Pi_\mathrm{min}} = Z\left(\frac{\Pi_0}{\Pi_\mathrm{min}^2}\right).
\end{equation}
 A local reduction in $\Pi$ relative to the background consequently allows regions of intrinsically low metallicities to exceed the clumping criterion. As such, we define a new normalisation parameter, $\chi$, as

\begin{equation}
    \chi = \frac{\Pi_\mathrm{min}^2}{\Pi_0},
\end{equation}

\noindent which captures both the local reduction in headwind and the resulting enhancement in metallicity induced by the pressure bump. When pressure variations are negligible, i.e. $\Pi_\mathrm{min} \approx \Pi_0$, then $\chi$ converges toward $\Pi_0$. Across all simulations, we find that $\chi$ correlates more tightly with the onset of strong clumping than $\Pi_\mathrm{min}$, $\Pi_0$, or $\tau_\mathrm{s}$ individually. For the bump amplitudes explored here, $\chi$ spans the range $0 \leq\chi \leq 0.05$ in our setups using $\Pi_0=0.05$ and $0.025 \leq \chi \leq 0.1$ in our setups using $\Pi_0=0.1$. Across our various setups, all simulations using $Z/\chi \geq 0.4$ exhibit the formation of dense particle filaments, whereas simulations using $Z/\chi \leq 0.2$ do not.  

Figure \ref{fig:clumpcrit} shows the outcome of the SI evolution for our simulations in the $Z/\chi$-$\tau_\mathrm{s}$ parameter space, alongside previously-established clumping thresholds (\citealp{Carrera15}; \citealp{Yang17}; \citealp{Li21}). Setups where $\Pi \rightarrow 0$ (i.e. strong pressure bumps) are omitted here, as $Z/\chi \rightarrow \infty$ in these cases. The clumping outcomes in our simulations closely follow the \cite{Yang17} criterion when using the $(Z/\chi)_\mathrm{crit}$ definition. This likeness may stem from our simulations also being long-lived, high-resolution, and two-dimensional evolutions of the pure SI, similar to the 2D configurations used in \cite{Yang17}. Regardless of initial metallicity and background pressure gradient, we infer that $(Z/\chi)_\mathrm{crit} \approx 0.3$ is the requirement to trigger strong clumping, based on our simulation results and the clumping criterion from \cite{Yang17}.

In our simulations, pressure bumps generate large enough increases in the azimuthal gas velocity to harbour extended regions where filament formation becomes highly active, with higher bump amplitudes ensuring that $Z/\chi$ is sufficiently large to sustain clumping-favourable conditions over a larger radial interval around $x(\Pi_{\rm{min}})$. This explains why setups well over $(Z/\chi)_\mathrm{crit}$ are also able to form dense filaments efficiently at distances far from $x(\Pi_\mathrm{min})$. For large bump amplitudes, drifting pebbles encounter a region of sufficiently increased metallicity and reduced headwind to exceed the $(Z/\chi)_\mathrm{crit}$ threshold long before they actually reach $x(\Pi_\mathrm{min})$. Interestingly, we only see strong clumping in the region where $x > x(\Pi_\mathrm{min})$ across all simulations, despite there not being any substantial differences in headwind on either side of $x(\Pi_\mathrm{min})$. This behaviour likely arises because, for $x > x(\Pi_\mathrm{min})$, the pebble drift steadily converges toward the most clumping-favourable point, whereas for $x < x(\Pi_\mathrm{min})$, it diverges from this point. Filaments also appear to disperse after they cross $x(\Pi_\mathrm{min})$, in tandem with the radial drift accelerating as the gradient in the azimuthal gas velocity becomes positive.   

The gradual increase in metallicity as the solid particle layer rearranges can also allow for the formation of filaments slightly further out than $x=0H$. The initial deceleration of the radial drift produces sufficiently strong particle density enhancements to trigger clumping around the bump center, with higher bump amplitudes incurring larger density enhancements here prior to the particle layer reaching a steady drift state. The innermost region of the bump also receives a progressively lower pebble flux once filaments form further out; therefore, the particle influx becomes less able to feed the innermost particle overdensities to an extent where they become radially stationary. Outermost filaments also tend to be the most likely to exceed the Roche density, as they eventually become the sole recipients of solids arriving from outside the bump. Nevertheless, we verified that the onset of clumping is not boosted by the pebble flux from outside the bump, as discussed in Appendix \ref{sec:robust}.   

Though the $(Z/\chi)_\mathrm{crit}$ definition of the clumping criterion manages to broadly predict filament formation from pure SI in simulations using both uniform and non-uniform pressure gradients, a more detailed investigation can be conducted into the role of the pressure bump width in creating clumping-favourable conditions. Strong particle concentration in setups with a uniform pressure gradient will generally continue to be amplified as filaments remain in a clumping-favourable region indefinitely, whilst clumping in pressure-variant regions is controlled by how fast the SI can grow relative to the time it takes for pebbles to cross the clumping-favourable region \citep{Carrera22}. We observe one case where further growth of the SI is hindered by filaments fully crossing the clumping-favourable region (see Figure \ref{fig:spacetimeplot5}), and we can postulate that narrowing the pressure bump could further inhibit the growth of the SI despite such a setup still exceeding the $(Z/\chi)_\mathrm{crit}$ threshold. Moreover, the prescription of the $(Z/\chi)_\mathrm{crit}$ criterion would likely break down if external turbulence were to be implemented, as injecting such turbulence would alter the mechanics of particle sedimentation beyond that which can be described by the pressure gradient parameter, $\Pi$, alone (\citealp{YoudinLithwick07}; \citealp{Li21}; \citealp{Eriksson26}).    

\section{Discussion}\label{discussion2}
\subsection{Stability of pressure bumps}\label{sec:bumpstability}
Our 2D simulations do not capture any non-axisymmetric hydrodynamical instabilities, the most notable acting on pressure bumps being the Rossby wave instability \citep{Lovelace99}. Narrow bumps of large amplitudes may induce large enough shear differences in the gas to break down the bump entirely. For an isothermal equation of state, the RWI has been found to occur when the epicyclic frequency, $\kappa$, is sufficiently reduced relative to the Keplerian orbital frequency (\citealp{Chang23}; \citealp{Chang24}), with the instability setting in once
\begin{equation}\label{eq:rwi}
   \kappa^2 \leq 0.6 \Omega_\mathrm{K}^2 . 
\end{equation}

We express the RWI threshold for our setups using the definition of the epicyclic frequency in a shearing box \citep{Chang24}. Taking into account our azimuthal gas velocity profile from Equation \ref{eq:vyg}, the substitution into Equation \ref{eq:rwi} yields  
\begin{equation}
   1-\frac{A}{1+A}\left( \frac{H}{w} \right)^2\leq 0.6,  
\end{equation}

\noindent which defines the condition under which the RWI is triggered at the point of maximum shear in our setups. The width and amplitudes we use ensure that this limit is not surpassed, with the highest bump amplitude used in our simulations yielding $\kappa^2/\Omega_\mathrm{K}^2 \simeq 0.78$. The RWI should therefore be a consideration for much narrower bumps ($w \ll H$) and higher bump amplitudes ($A \gg 0.1$) than ours.

Additionally, deformation of the pressure bump occurs from the back-reaction between solid particles and gas, as the bump is exempt from any reinforcement mechanism. This results in partial damping of the pressure bump as gas is advected outward, significantly altering the radial gas density profile relative to the background and partially deforming the azimuthal gas velocity profile. The damping of the azimuthal gas velocity is very minimal across all simulations, with most of the deformation stemming from gas being accelerated azimuthally by solid particles. In essence, deformation of the azimuthal gas velocity occurs in tandem with the SI becoming active, thus likely being a byproduct of the formation of dense particle filaments. Moreover, planetesimal formation would likely be thwarted if particle back-reaction were to be disregarded, despite particles being accumulated at the pressure bump \citep{Xu22}. Extensive reinforcement of the pressure bump may also inhibit the non-linear growth of the SI by undoing modifications in the relative gas-particle velocities, so bump deformation must consequently be allowed for the gas–particle feedback to evolve naturally. Further discussion on the topic of pressure bump deformation is provided in Appendix \ref{sec:deformapp}.

\subsection{Resolution considerations}
A reduction in pressure support causes the characteristic radial distance between the fastest growing modes of the SI to shrink. This becomes a numerical issue when simulating bumps of Keplerian and super-Keplerian amplitudes, as the resonant wavelength of the SI is directly related to the pressure support. As the pressure gradient approaches zero, the streaming instability nears an infinitesimally small, and thus unresolvable, radial wavelength, especially for particles of lower Stokes numbers \citep{Squire20}. As our grid resolution is very high, with every grid cell having a physical size of $\Delta x \approx 7.8\cdot 10^{-4} H$, we avoid the issue of resolving the smallest possible wavelengths for the majority of our simulations. The outliers are the $A=0.14$ and $A=0.17$ runs with $\Pi_0 = 0.05$, constituting bumps where $\Pi = 0$ at one point for $A=0.14$ and two points for $A=0.17$. The grid is unable to resolve the fastest growing modes of the SI in small regions around points where $\Pi = 0$ for these simulations, though mainly for Stokes numbers $\tau_\mathrm{s} = 0.01$ and $\tau_\mathrm{s}=0.03$. However, as these regions are extremely narrow thanks to the high grid resolution, the fastest growing modes are only unresolvable over a small radial interval around $x(\Pi_\mathrm{min})$. The largest unresolvable region across our setups is obtained for a background pressure parameter of $\Pi_0=0.05$, a bump amplitude of $A=0.14$, and particles of Stokes number $\tau_\mathrm{s}=0.01$, given that the resonant wavelength of the fastest growing SI modes decreases with both $\Pi$ and $\tau_\mathrm{s}$ \citep{Squire20}. This bump setup hosts a radial interval of roughly $\pm0.3H$ around $x(\Pi_\mathrm{min})$ where $\lambda_\mathrm{SI,x}$ is smaller than $\Delta x$. As discussed in Sections \ref{sec:innerdisc} and \ref{sec:outerdisc}, simulations using high bump amplitudes also exhibit highly resilient filament formation in the pressure bump, with dense filaments most often emerging at locations outside the unresolvable region (see Figure \ref{fig:rhop_matrix}). Furthermore, the growth rate of the SI nears zero for very small pressure gradients due to the minimal particle drift (\citealp{Youdin05}; \citealp{Abod19}; \citealp{Squire20}). Negligible drift thus causes regions near $\Pi=0$ to be virtually incapable of producing azimuthal particle filaments via the SI, regardless of grid resolution.  

\section{Conclusions and future work}\label{conclusion}

We tested the influence of weak pressure bumps on triggering planetesimal formation via the streaming instability. Our main finding is that fairly minor, non-reinforced pressure perturbations facilitate the formation of particle filaments by the streaming instability. Our key takeaways include:

\begin{enumerate}

\item Inner regions in the disc, which are generally less affected by pressure support, are able to form dense particle filaments at pressure bumps of very low density amplitudes. Amplitudes down to $A=0.04$, corresponding to a reduction of the pressure gradient parameter from $\Pi_0=0.05$ to $\Pi_{\rm{min}}=0.035$, are sufficient for planetesimal formation for pebbles in the millimetre-to-centimetre size range. \\

\item Outer regions in the disc, which experience a larger degree of pressure support, are still able to form dense particle filaments at weak pressure bumps for pebbles in the millimetre-to-centimetre size range, though they require a larger density amplitude ($A\geq 0.14$, corresponding to a reduction of the pressure gradient parameter from $\Pi_0=0.1$ to $\Pi_{\rm{min}}\leq 0.05$) to trigger strong clumping compared to simulations using a lower background pressure 
support. \\

\item Planetesimal formation is promoted at pressure bumps in regions of higher metallicities ($Z=0.02$), with an increase in metallicity requiring lower bump amplitudes to harbour conditions favourable to strong clumping. \\

\item The clumping criterion at pressure bumps can be broadly described by factoring in the differences in the radial advection of pebbles in a region where $\Pi$ varies. The corresponding normalised metallicity can be prescribed by the definition $Z/\chi$, where $\chi = \Pi_\mathrm{min}^2/\Pi_0$. $\Pi_\mathrm{min}$ decreases with higher bump 
amplitude, $A$. We find $(Z/\chi)_\mathrm{crit} \approx 0.3$ to match the intervals in the $Z/\chi$-$\tau_\mathrm{s}$ parameter space where solid particle clumping should and should not take place. This threshold also agrees with the \cite{Yang17} clumping criterion that did not include any pressure bump. \\

\item Whilst pressure bumps of high amplitudes run the risk of becoming Rossby wave unstable, the discovery that weak pressure bumps can easily facilitate planetesimal formation from the streaming instability also largely circumvents this issue. In addition, weak pressure bumps remain relatively robust against deformation through back-reaction between pebbles and gas.

\end{enumerate}

In conclusion, weak pressure bumps instigated by non-reinforced gas perturbations, such as those emerging from inverse cascades in magnetohydrodynamical turbulence (\citealp{Johansen09}; \citealp{Bethune17}; \citealp{Riols19}; \citealp{Eriksson26}), are strong candidates for sites of planetesimal formation by the streaming instability in protoplanetary discs. This finding provides an explanation of how planetesimals may form without the need for large-scale gaps carved by planets or any high-amplitude pressure bump of unspecified origin that may be unstable to the Rossby wave instability. 

Prospects for future work on this topic include expanding simulations into 3D. This may reveal that even smaller metallicity enhancements may be sufficient to trigger planetesimal formation driven by the streaming instability at pressure bumps compared to those seen in our simulations (\citealp{Li21}; \citealp{Lim26}), while also allowing for the evolution of self-gravity and thus the full collapse of particle clumps into planetesimals (\citealp{Simon16}; \citealp{Schafer17}; \citealp{Li19}). It will similarly be important to examine how the radial width of pressure bumps may impact the growth of the streaming instability within the bump, as well as how dust evolution interacts with the streaming instability in particle pile-ups at pressure bumps. In the latter, the effects of coagulation and fragmentation of individual pebbles must be considered, which requires a polydisperse approach (\citealp{Yang21}; \citealp{Zhu21}; \citealp{Ho24}; \citealp{Carrera25}; \citealp{Vallucci-Goy26}). Finally, young discs have significant density variations, but are also subject to disturbances from streamers and non-radial infall that may drive macro-turbulence in the disc (\citealp{Kuffmeier18}; \citealp{Zhao25}; \citealp{Holm26}). The impact of long-wavelength, non-radial disturbances on the ability of pressure bumps to drive planetesimal formation is thus also an important future question. Overall, the results of this work provide the first step in understanding how weak pressure maxima influence the streaming instability. The large parameter space we explore allows our results to be interpreted with generality, and further parameter-specific investigations will inevitably be useful in quantifying the full role of pressure bumps in the early formation stage of protoplanets.

\begin{acknowledgements}
       A.J.~acknowledges funding from the Carlsberg Foundation (Semper Ardens: Advance grant FIRSTATMO),  T.H.~ acknowledges funding from the Independent Research Fund Denmark through grant No. DFF 10.46540/4283-00305B. A.J.~is grateful for  discussions with Daniel Carrera on  different approaches to setting up gas pressure bumps. The Tycho supercomputer at the University of Copenhagen SCIENCE HPC center was used for carrying out the simulations.
\end{acknowledgements}

\appendix
\nolinenumbers
\section{Quantifying bump deformation}\label{sec:deformapp}

    \begin{figure}[t!]
   \centering
   \includegraphics[width=\hsize]{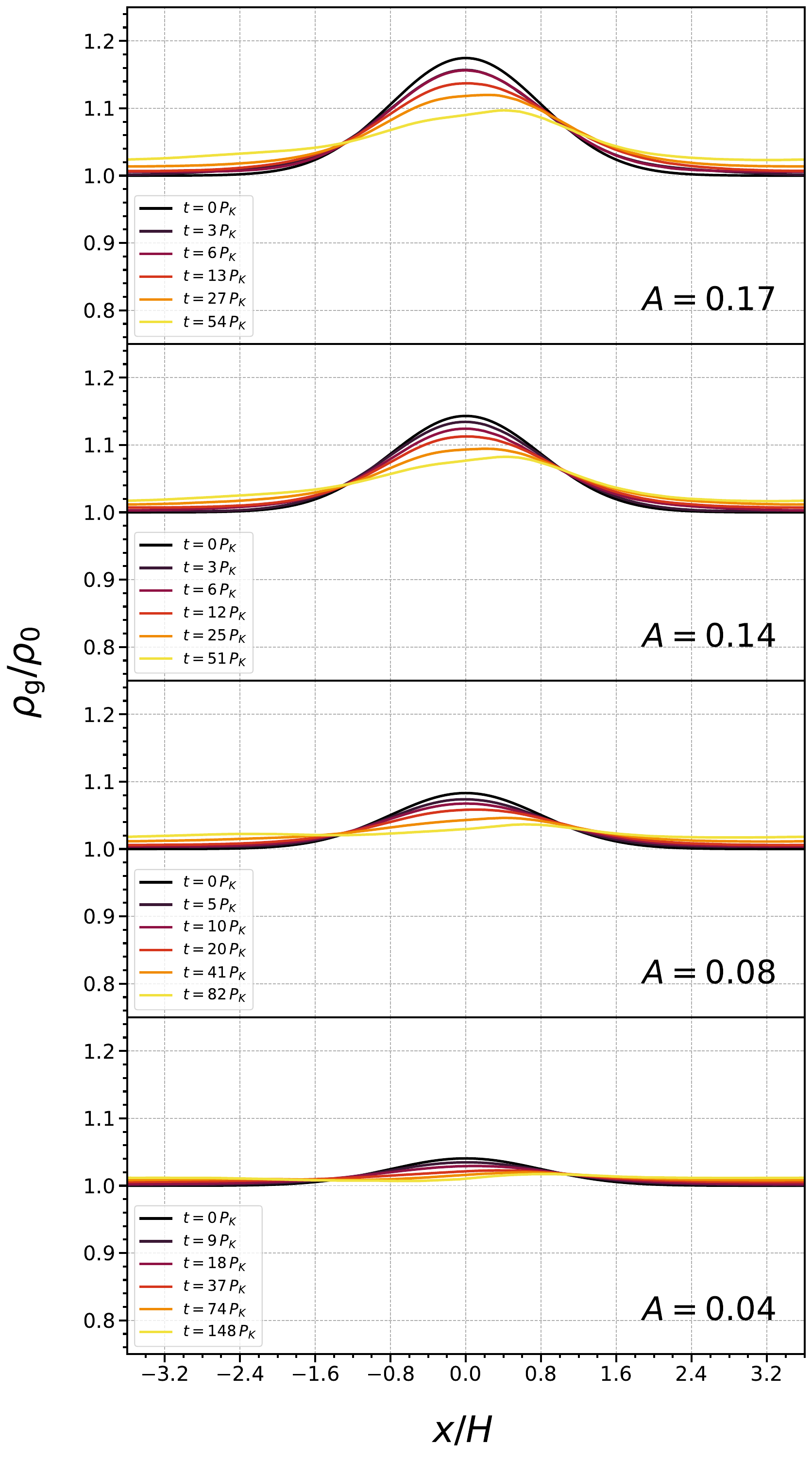}
      \caption{Evolution of the gas density profiles relative to the background at different bump amplitudes, for particles of Stokes number $\tau_\mathrm{s}=0.03$, a background pressure gradient parameter of $\Pi_0=0.05$, and an initial metallicity of $Z=0.01$.}
         \label{fig:deform2}
   \end{figure}

  \begin{figure}[t!]
   \centering
   \includegraphics[width=\hsize]{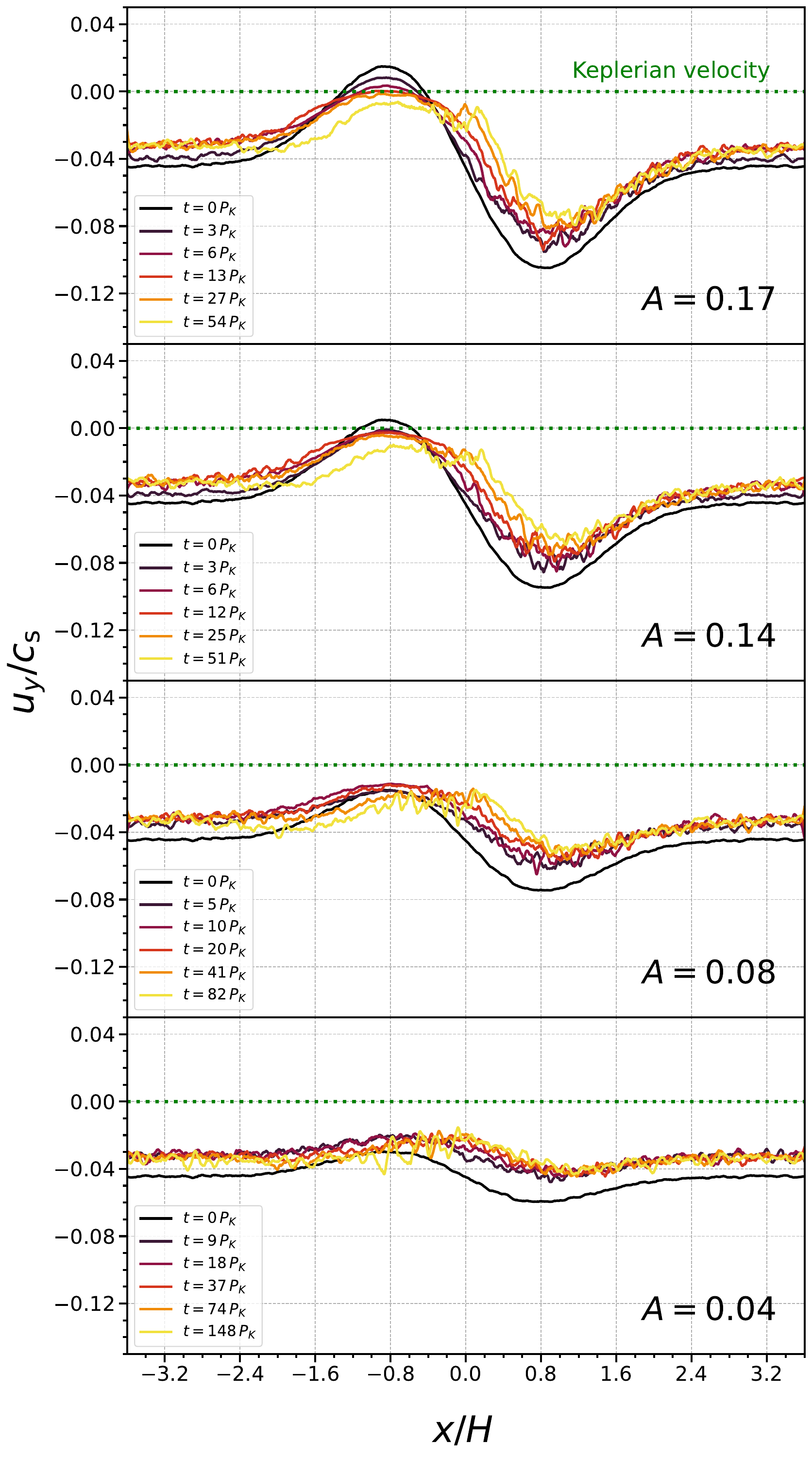}
      \caption{Evolution of the Keplerian-shear-subtracted azimuthal gas velocity profiles at different bump amplitudes for particles of Stokes number $\tau_\mathrm{s}=0.03$, background pressure gradient parameter $\Pi_0=0.05$, and metallicity $Z=0.01$. Note that the azimuthal gas velocity profiles are slightly offset from their analytical profiles at $t=0 P_\mathrm{K}$, as particle back-reaction is active from the very beginning of each simulation.}
         \label{fig:deform}
   \end{figure}

The deformation of the pressure bump through interaction with solid particles can be further understood by recognising how friction forces between gas and solids alter the azimuthal gas velocity profile, providing an extension to our discussion from Section \ref{sec:bumpstability}. The two-way back-reaction between solid particles and gas modifies the gas velocity by inducing an acceleration, 
\begin{equation}
    \mathbf{a}_\mathrm{g} = -\frac{\epsilon}{t_\mathrm{stop}}[\mathbf{u}-\mathbf{w}],
\end{equation}
where $\mathbf{u}=(u_x, u_y, u_z)$ are the Keplerian-shear-subtracted velocities of gas and $\mathbf{w}=(w_x, w_y, w_z)$ are the Keplerian-shear-subtracted velocities of solid particles. As such, the azimuthal gas velocity profile induced by a pressure bump may be deformed by friction from the solids, depending on the solid-to-gas ratio, $\epsilon$, and the particle stopping time, $t_{\mathrm{stop}}$. Deformation via friction forces alters the bump in the gas density relative to the background as gas is advected outward (see Figure \ref{fig:deform2}), though this is not necessarily indicative of extensive deformation of the azimuthal gas velocity profile. The peak in $u_y$ for bumps of higher amplitude tends to decrease slightly due to the back-reaction, whilst the peak in $u_y$ for lower-amplitude bumps increases in tandem with the rest of the azimuthal gas velocity profile being accelerated by interaction with solid particles (see Figure \ref{fig:deform}). The largest deformation tends to occur around $x=0H$, which is the inflection point where solids briefly accumulate before reaching a steady drift state. In general, increases in the azimuthal gas velocity are seen at intervals where the solid-to-gas ratio is large, coinciding with regions where particle filaments are produced. Similarly, regions that become solid-deficient, such as those starward of where dense particle filaments are formed, see a slight decrease in azimuthal gas velocity. Damping of the azimuthal gas velocity peak does not occur in the absence of particle filaments for sub-Keplerian bumps, instead occurring in conjunction with the formation of dense filaments. Super-Keplerian bumps also become damped towards Keplerian speeds, as solid particles instil a drag force opposite the direction of the gas flow in the region where $\Pi<0$. The decrease in particle density on the starward side of dense filaments, and the anti‑parallel drag exerted by solids on gas bumps with slightly super‑Keplerian peak velocities, are the two distinct sources of damping in our simulations. As a result, we can strongly argue that any deformation of the pressure bump does not inhibit the formation of dense particle filaments, and that it is rather an indication of filament formation in most cases.

\section{Extended pressure bumps}\label{sec:robust}

    \begin{figure}[t!]
   \centering
   \includegraphics[width=\hsize]{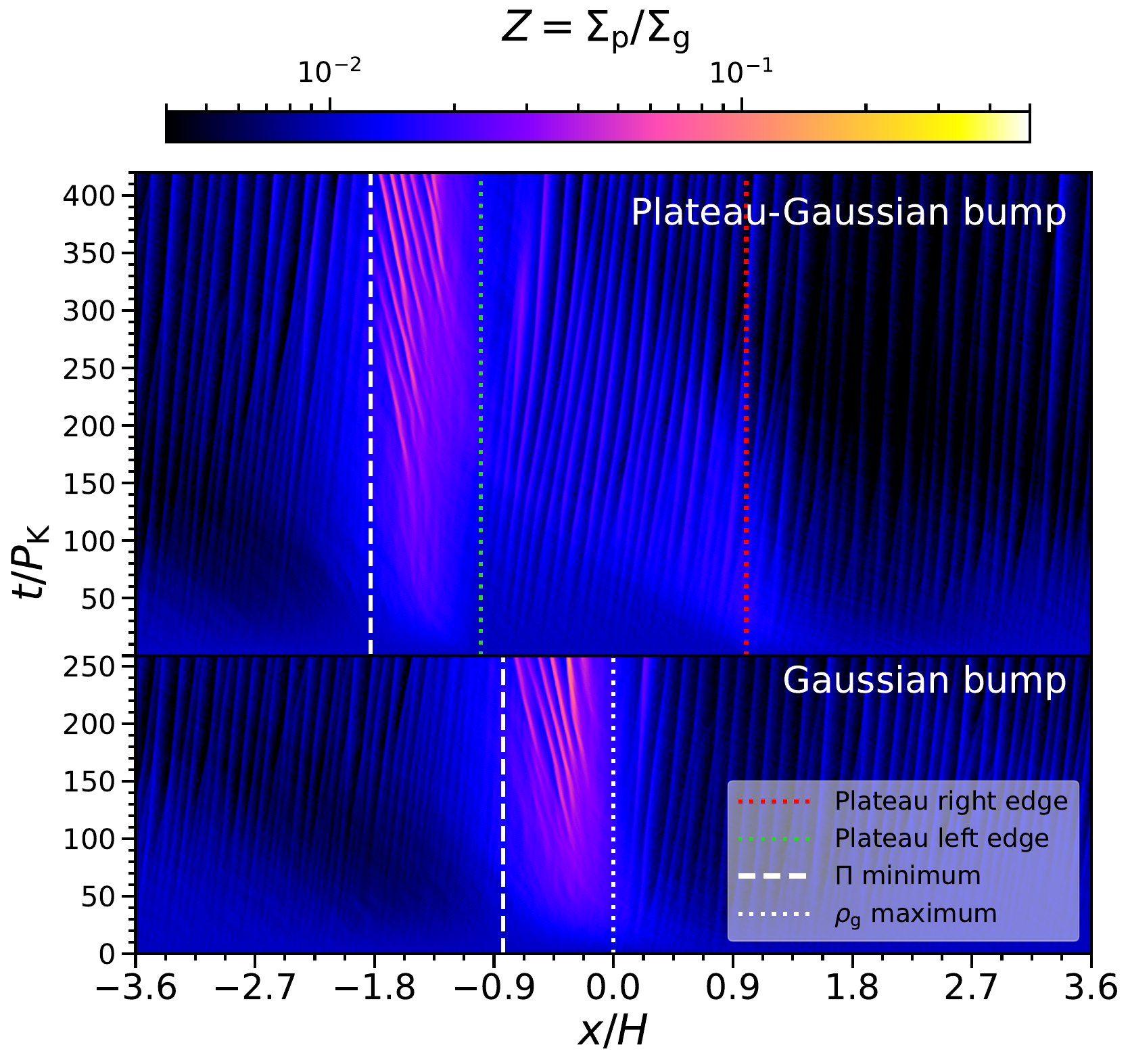}
      \caption{Time evolution of the metallicity, $Z=\Sigma_\mathrm{p}/\Sigma_\mathrm{g}$, using a Gaussian bump setup (bottom) and a plateau-Gaussian bump setup (top). Both cases use solid particles of Stokes number $\tau_\mathrm{s} = 0.01$, a background pressure parameter of $\Pi_0 = 0.1$, an initial metallicity of $Z=0.01$, and a pressure bump amplitude of $A=0.14$. Both panels show the full radial domain of each simulation.}
         \label{fig:sg001}
   \end{figure}

       \begin{figure}[t!]
   \centering
   \includegraphics[width=\hsize]{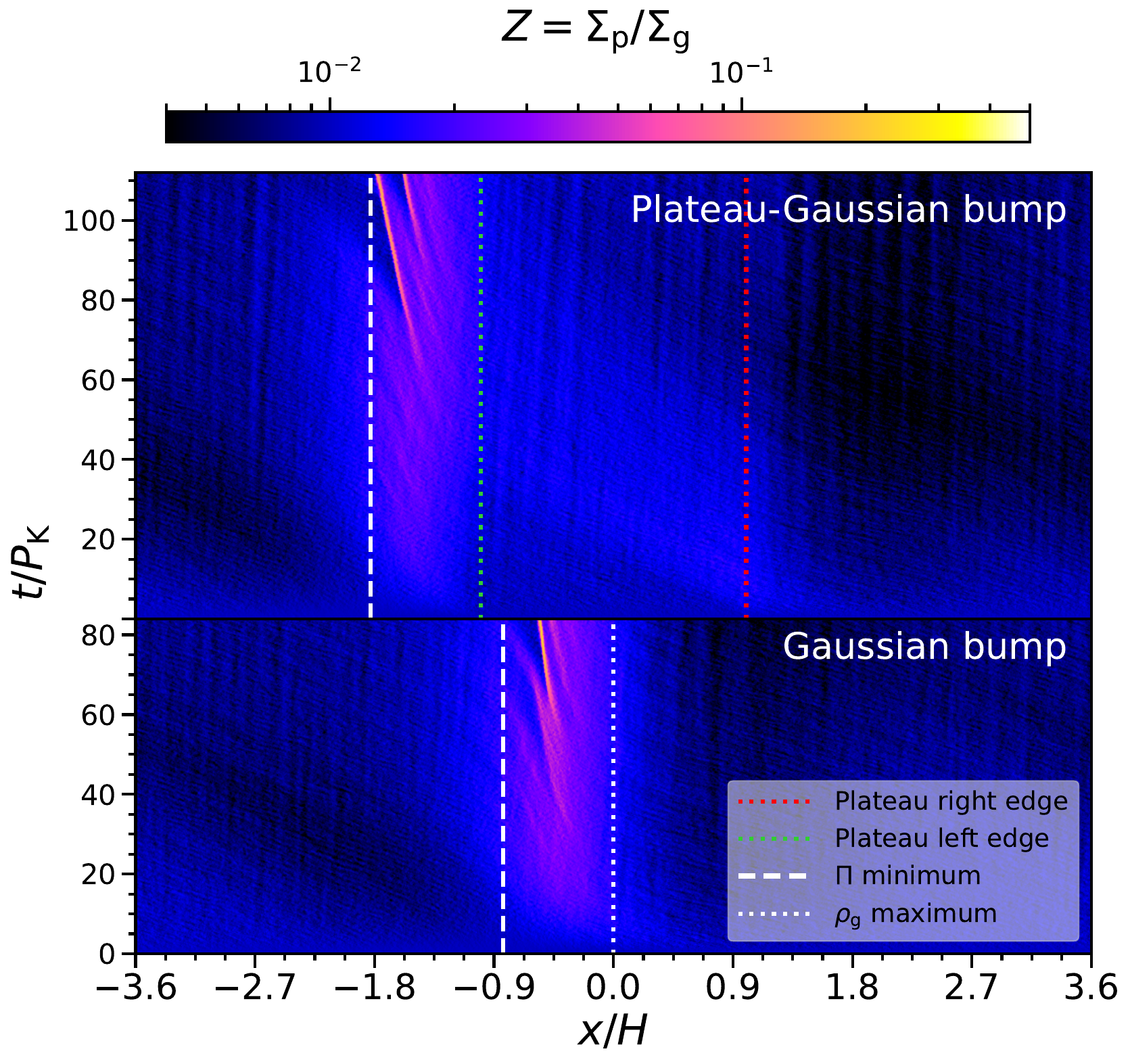}
      \caption{Similar to Figure \ref{fig:sg001}, instead using particles of Stokes number $\tau_\mathrm{s}=0.03$.}
         \label{fig:sg003}
   \end{figure}

  \begin{figure}[t!]
   \centering
   \includegraphics[width=\hsize]{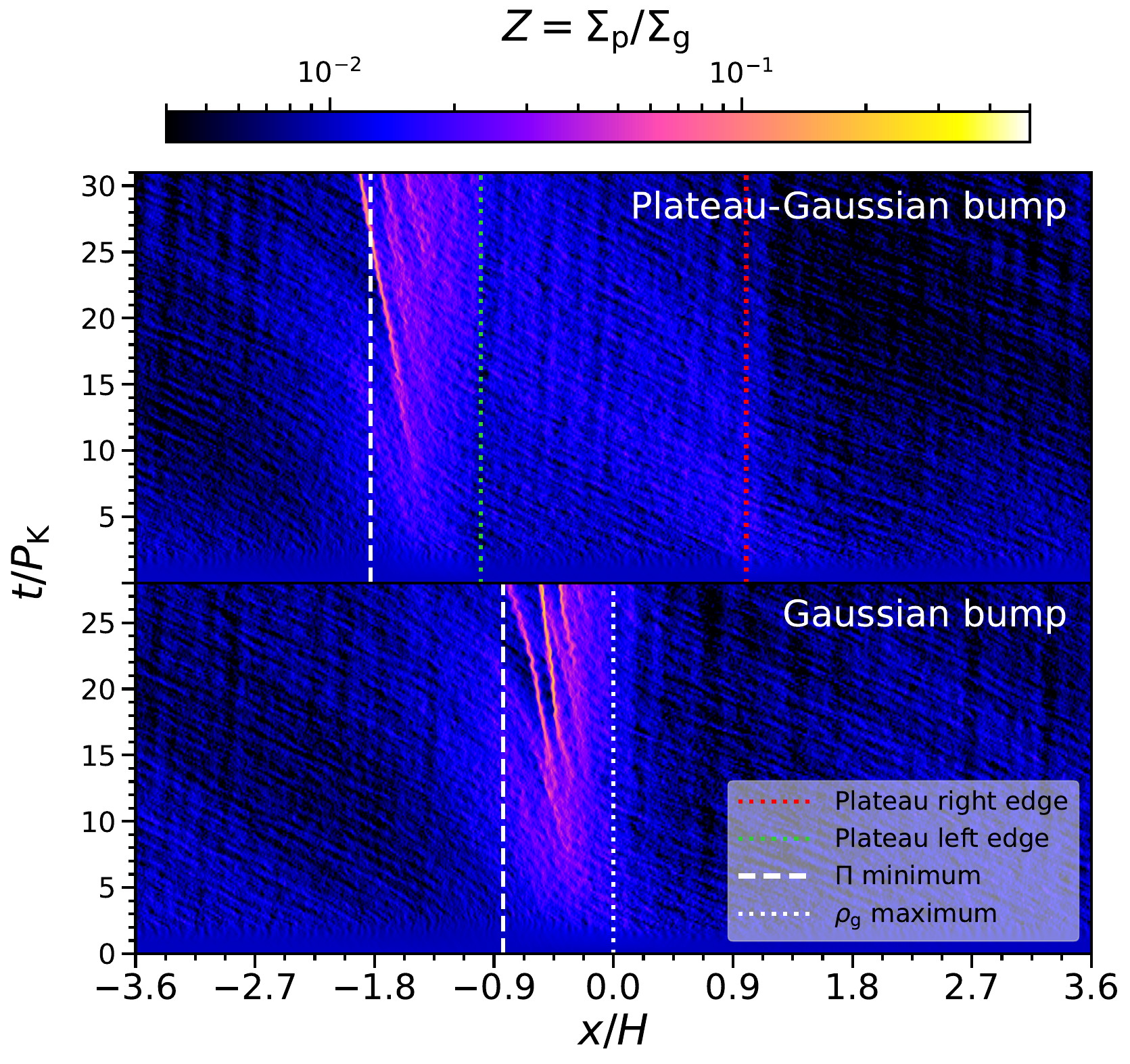}
      \caption{Similar to Figures \ref{fig:sg001} and \ref{fig:sg003}, instead using particles of Stokes number $\tau_\mathrm{s}=0.1$.}
         \label{fig:sg01}
   \end{figure}

The starward side of a pressure bump appears to be the key component in promoting non-linear growth of the SI, as this constitutes the location of the steepest reduction in pressure support and the largest increase in particle density relative to the gas. In general, we find that the conditions for strong clumping rely very sensitively on the minimum headwind, $\Pi_\mathrm{min}$, though filaments can form preferentially in regions of heightened pebble flux. As such, we must ensure that the formation of particle filaments around the location of the minimum azimuthal headwind is a general feature of pressure bumps and not simply a side effect of our particular setup. We therefore test whether filament formation depends definitively on the steepness of the bump's shoulders, rather than on the pebble flux from outside the bump, by isolating the left and right sides of the azimuthal gas velocity profile. This is done by using a setup with a radially-extended ''plateau-Gaussian'' bump, where $ \rho_\mathrm{g}(x,z)$ is defined by

\begin{equation}
 \rho_\mathrm{g}(x,z) = 
\begin{cases}
    \rho_0\left[1+A \mathrm{e}^{-(x+1)^2/2w^2} \right]\mathrm{e}^{-z^2/2H^2}& -3.6H \leq x < -H, \\
    \rho_0\left[1+A\right]\mathrm{e}^{-z^2/2H^2} & -H \leq x \leq H, \\
    \rho_0\left[1+A \mathrm{e}^{-(x-1)^2/2w^2} \right]\mathrm{e}^{-z^2/2H^2} & H < x \leq 3.6H.
\end{cases}
\end{equation}

Geostrophic balance will hereby instil an azimuthal gas velocity similar to that of Equation \ref{eq:vyg}, but where the $u_y$ maximum now lies at $x(\Pi_\mathrm{min}) \approx -1.8H$ and the $u_y$ minimum lies at $x(\Pi_\mathrm{max}) \approx 1.8H$. The peak and trough of the azimuthal gas velocity profile are now separated by a plateau. Between the left and right edges of the plateau, situated at $x=\pm H$, $u_y$ takes on the constant background value. We compare the evolution of the SI in this setup to that of previous runs using $\Pi_0=0.1$, $Z=0.01$, and $A=0.14$, for Stokes numbers $\tau_\mathrm{s}=0.01$, $\tau_\mathrm{s}=0.03$, and $\tau_\mathrm{s}=0.1$. Individual comparisons between the outcomes of the Gaussian and plateau-Gaussian pressure bump setups for all three tested Stokes numbers are shown in Figures \ref{fig:sg001}, \ref{fig:sg003}, and \ref{fig:sg01}. In each of the plateau-Gaussian setups, filaments are formed on the left shoulder of the pressure bump, specifically in the interval where $\Pi_0 <\Pi \leq \Pi_\mathrm{min}$. This is in accordance with the filament formation seen in the equivalent Gaussian bump simulations for each of the three Stokes numbers used. As the azimuthal gas velocity profile in the interval $-0.8 H < x < 0 H$ for the Gaussian bump is identical to the interval $-1.8 H < x < -H$ for the plateau-Gaussian bump, it is unsurprising that the region favourable to non-linear SI growth appears to be shifted starward by exactly one gas scale height in the plateau-Gaussian bump setup. The timescales at which filaments are formed in the plateau-Gaussian bump setups are comparable to those of the Gaussian bump setups, though filaments initially form slightly closer to $\Pi_\mathrm{min}$ in the plateau-Gaussian bump setups. The latter is likely due to solid particles being redistributed slightly differently before a steady drift state comes into effect, given the differences in the radial profiles of the inward drift speed of solid particles when compared to the Gaussian bump setups. In any case, this robustness test strengthens the argument that filament formation from the SI in a pressure bump is generally governed by the reduced pressure support, rather than the pebble flux from outside the bump.

\end{document}